\documentclass[final,5p,times,twocolumn]{elsarticle}

\usepackage{amssymb}
\usepackage{amsthm}
\usepackage{amsmath}
\usepackage{braket}
\usepackage{graphicx}
\allowdisplaybreaks

\usepackage{listings}
\usepackage{xcolor}
\usepackage{upquote}

\usepackage{hyperref}
\hypersetup{
	colorlinks=true, 
	linktoc=all,     
	linkcolor=blue,  
}

\newcounter{bla}

\journal{Computer Physics Communications}
\begin{document}

\begin{frontmatter}

\title{{\tt Qcombo}: A Python Package for Automated Commutator Calculations of Quantum Many-Body Operators}
 
\author[a,b]{L. H. Chen} \ead{chenlh73@mail2.sysu.edu.cn} 
\author[a,b]{Y. Li}\ead{liyi226@mail2.sysu.edu.cn}
\author[c,d]{H. Hergert}\ead{hergert@frib.msu.edu}
\author[a,b]{J. M. Yao   \corref{author}}\ead{yaojm8@sysu.edu.cn}

\cortext[author] {Corresponding author}
\address[a]{School of Physics and Astronomy, Sun Yat-sen University, Zhuhai 519082, P.R. China}
\address[b]{Guangdong Provincial Key Laboratory of Quantum Metrology and Sensing, Sun Yat-Sen University, Zhuhai 519082, P.R. China }

\address[c]{Facility for Rare Isotope Beams, Michigan State University, East Lansing, Michigan 48824-1321, USA }
\address[d]{Department of Physics \& Astronomy, Michigan State University, East Lansing, Michigan 48824-1321, USA }

\begin{abstract}

{\tt qcombo} is a Python package for the symbolic evaluation of commutators between general quantum many-body operators expressed in normal-ordered form using the generalized Wick theorem. The package provides an automated and systematic framework for generating the corresponding algebraic expressions, significantly reducing the risk of human error in lengthy and complex analytical derivations. It is designed to assist the development and implementation of modern many-body methods in nuclear physics, quantum chemistry, and related fields.  The functionality and workflow of the package are demonstrated through an application to the in-medium similarity renormalization group (IMSRG) method, which has been widely used for nuclear ab initio calculations. As a representative example, {\tt qcombo} is employed to automatically generate the complete set of multi-reference IMSRG flow equations with operators truncated at the normal-ordered three-body level.

   \end{abstract}
   

\end{frontmatter}

\noindent \textbf{PROGRAM SUMMARY}

\begin{small}
\noindent
{\em Program Title:}    {\tt qcombo }                                      \\
{\em CPC Library link to program files:} (to be added by Technical Editor) \\
{\em Developer's repository link:}  https://github.com/chenlh73/qcombo.git \\
{\em Licensing provisions:}  MIT \\
{\em Programming language:}    Python (Requires-Python >=3.12)   \\
{\em Required additional packages:}    SymPy                              \\
{\em Supplementary material:}   https://pypi.org/project/qcombo/                              \\
{\em Nature of problem:} 
  Many modern quantum many-body methods require the evaluation of commutators between general many-body operators expressed in normal-ordered form.   The derivation of these commutators involves repeated applications of the generalized Wick theorem and typically generates a large number of algebraic terms with complicated index structures. Carrying out these derivations manually is therefore extremely labor-intensive and prone to human error, especially when higher-body operators or multi-reference states are considered. \\
{\em Solution method:}  The program implements a symbolic framework based on the generalized Wick theorem to automatically evaluate commutators between normal-ordered many-body operators. It systematically generates all operator contractions and constructs the resulting expressions, which are then simplified using symmetry properties of operators and by transforming to the natural-orbital basis. The final results are provided in symbolic form for further analytical or numerical use.\\
{\em Additional comments including restrictions and unusual features:} None.\\

\end{small}

\allowdisplaybreaks


\section{Introduction}
\label{Introduction}

The quantum many-body problem remains one of the central challenges in modern physics, as it underpins our understanding of atomic nuclei, condensed matter systems, quantum chemistry, and strongly correlated materials. Its solution is notoriously difficult because the Hilbert space grows exponentially with particle number and degrees of freedom, rendering exact treatments feasible only for the smallest systems.

Over the past decades, a variety of systematically improvable methods have been developed to address this problem at manageable computational cost. Among them are approaches based on the coupled-cluster (CC) theory \cite{Coester:1960,Bartlett:2007,Hagen:2014} and similarity renormalization group (SRG)~\cite{Wilson:1994,Wegner:1994}, including the in-medium SRG (IMSRG) in nuclear physics~\cite{Tsukiyama:2011PRL} and the driven SRG in quantum chemistry~\cite{Francesco:2014}. In these frameworks, the time and memory requirements scale polynomially with the size of the underlying single-particle basis and depend only indirectly on the particle number. However, their practical implementation relies fundamentally on the evaluation of commutators between many-body operators, which constitutes a central computational task.  

In the IMSRG, operators are normal ordered with respect to a chosen reference state, which encodes physical information of the targeted nuclear system, and then truncated at a fixed $k-$body rank. The baseline for most applications is the so-called normal-ordered two-body (NO2B) approximation, which achieves a balance between accuracy and computational cost~\cite{Hergert:2016PR,Stroberg:2019ARNPS}. A common choice for the  reference state is a Slater determinant, leading to the single-reference SR-IMSRG(2) implementation, which generally provides an adequate description for closed-shell systems, i.e., systems with a pronounced excitation gap. For open-shell systems, collective correlations become increasingly important, and the approximation becomes less reliable due to the neglect of higher-body contributions. Improving the accuracy therefore requires the inclusion of higher-body terms, as in SR-IMSRG(3)~\cite{Heinz:2021,Stroberg:2024}, where all operators are truncated at the normal-ordered three-body (NO3B) level. Yet such an extension entails a substantial increase in computational cost: SR-IMSRG(3) scales as $\mathcal{O}(N^9)$, where $N$ denotes the number of single-particle states, compared with the $\mathcal{O}(N^6)$ scaling of SR-IMSRG(2). This computational cost makes the full IMSRG(3) challenging for applications to heavy nuclei. Consequently, several approximate IMSRG(3) schemes that partially retain higher-body correlations have been proposed~\cite{Heinz:2021,Stroberg:2024,He:2024}.

In the IMSRG framework, one has the freedom to decide which correlations are captured by the IMSRG transformation, and which by the reference state. Thus, a possible alternative strategy to using a high SR-IMSRG($k$) truncation is the use of a correlated reference state, leading to the multi-reference (MR) framework~\cite{Hergert:2013PRL,Gebrerufael:2017PRL,Francesco:2018,Hergert:2017PS,Yao:2020PRL}. In this approach, part of the higher-order particle–hole correlations is incorporated directly into the reference state. Nevertheless, MR-IMSRG(2) still suffers from truncation errors due to the omission of genuine three-body and higher-rank contributions, which is indicated by the moderate dependence of the results on the different choices of reference states~\cite{Zhou:2025}. To further improve accuracy and to systematically assess truncation effects, which is important for applications to nuclear structure phenomena or the computation of nuclear inputs for the new physics searches in high-precision frontiers~\cite{Belley:2024PRL,Hayen:2024}, one must proceed to higher truncation levels, such as MR-IMSRG(3). As the operator rank increases, the number of terms appearing in the commutators grows rapidly, rendering manual derivations both tedious and prone to error.

To tackle the complexity of high-rank commutators in advanced quantum many-body methods, such as MR-IMSRG(3) and MR-CCSDT with full triples truncations, we present the {\tt qcombo} Python package for the symbolic evaluation of many-body operator commutators based on {\tt SymPy}. The package automates the generation and algebraic simplification of commutators involving arbitrary normal-ordered many-body operators in the $M$-scheme. Furthermore, it interfaces with the existing {\tt AMC} package~\cite{Tichai2020amc}, enabling the direct export of the resulting expressions in the angular-momentum-coupled ($J$-scheme) representation required for practical nuclear structure calculations.  

We note that, unlike the Python package {\tt sympy.physics.secondquant}, which is formulated with respect to the particle-vacuum reference state, {\tt qcombo} is designed for an arbitrary reference state, covering both single-reference and multi-reference cases. This enables the automatic derivation and simplification of operator expressions within the framework of a generalized normal ordering \cite{Kutzelnigg1997}. In the vacuum limit, where all density matrices vanish, the expressions generated by {\tt qcombo} reduce to the corresponding results obtained in vacuum-normal-ordered formalisms. This feature makes {\tt qcombo} particularly suitable for applications in multi-reference many-body methods such as MR-CC and MR-IMSRG.  It is also worth noting that several symbolic-algebra frameworks have been developed to automate the derivation and implementation of quantum many-body theories in quantum chemistry. These packages provide powerful capabilities for symbolic tensor manipulation, automatic code generation, and scalable numerical implementation. The \texttt{qcombo} complements these efforts, with particular attention to the specific needs of nuclear many-body theory.  A more detailed comparison of {\tt qcombo} with these frameworks is provided in ~\ref{sec:comparison_quantum_chemistry}.

This paper is organized as follows. In Sec.~\ref{sec:Theoretical Framework}, we outline the theoretical foundation of the IMSRG and summarize the commutation relations of normal-ordered many-body operators. The design principles and usage of the {\tt qcombo} package are described and illustrated with examples in Sec.~\ref{sec:Package_Description}. Benchmark calculations and applications to MR-IMSRG(3) are presented in Sec.~\ref{sec:Application}. Conclusions and outlook are given in Sec.~\ref{sec:Conclusion}.

\section{Theoretical Framework}
\label{sec:Theoretical Framework}

The basic idea of the IMSRG method is to decouple the off-diagonal elements of a Hamiltonian in the configuration space by introducing a set of continuous unitary transformations~\cite{Hergert:2016PR} 
\begin{equation}
    H(s) \equiv U(s) H(0) U^\dagger(s).
\end{equation}
Here, $H(0)$ is the initial Hamiltonian. The IMSRG flow equation can be derived by differentiating the transformed Hamiltonian $H(s)$ with respect to the flow parameter $s$,
\begin{equation}
\label{eq:SRG_flow_equation}
    \frac{d}{ds} H(s) = [\eta(s), H(s)],
\end{equation}
where the generator $\eta(s)$ is defined by
\begin{equation}
    \eta(s) \equiv \frac{dU(s)}{ds} U^\dagger(s) = -\eta^\dagger(s). 
\end{equation}
In practical applications, the unitary transformation $U(s)$ is determined by specifying a particular generator $\eta(s)$ instead \cite{Hergert:2016PR,Hergert:2017PS}. 

The Hamiltonian, including up to three-body interactions, can be expressed in second-quantized form 
\begin{equation}
    H = \sum_{ij} t^{i}_{j} A^{i}_{j} + \frac{1}{4}\sum_{ijkl} v^{ij}_{kl} A^{ij}_{kl} + \frac{1}{36} \sum_{ijklmn} w^{ijk}_{lmn} A^{ijk}_{lmn}.
\end{equation}
Here $t^{i}_{j}$ denotes the matrix elements of the one-body kinetic-energy term, while $v^{ij}_{kl}$ and $w^{ijk}_{lmn}$ represent the antisymmetrized matrix elements of the two-body and three-body terms, respectively. The notation $A^{i_1\dots i_k}_{j_1 \dots j_k}$ compactly represents a string of fermionic creation and annihilation operators:
\begin{equation}
A^{i_1\dots i_k}_{j_1 \dots j_k} = a^{\dagger}_{i_1} \dots a^{\dagger}_{i_k} a_{j_k} \dots a_{j_1},
\label{eq:operator_string}
\end{equation}
where the operators $a^\dagger_i$ and $a_i$ respectively create and annihilate a particle in the single-particle state $|i\rangle$. 

To solve the IMSRG flow equation, we need to calculate the commutators of the generator and Hamiltonian, both of which are many-body operators and are normal-ordered with respect to an arbitrary reference state $\ket{\Phi}$. The one-body normal-ordered operator is defined as
\begin{equation}
    \{A^i_j\} = A^i_j - \rho^i_j,
\end{equation}
where $\rho^i_j$ denotes the one-body density matrix of $\ket{\Phi}$. More generally, the normal-ordered $k$-body operator can be defined recursively~\cite{Kutzelnigg1997} as
\begin{equation}
\begin{aligned}
    \{A\}^{(k)} &\equiv \{A^{i_1i_2\dots i_k}_{j_1j_2 \dots j_k}\} \\
    &= A^{i_1\dots i_k}_{j_1 \dots j_k} -{\cal A}[\rho^{i_1}_{j_1}\{A^{i_2\dots i_k}_{j_2 \dots j_k}\}] \\
    & - {\cal A}[\rho^{i_1 i_2}_{j_1 j_2} \{A^{i_3\dots i_k}_{j_3 \dots j_k} \}] 
    - \dots - \rho^{i_1i_2\dots i_k}_{j_1 j_2\dots j_k},
\end{aligned}
\end{equation}
where  the antisymmetrization operator ${\cal A}$ generates all possible unique permutations of upper indices and lower indices, and $\rho^{i_1\dots i_k}_{j_1 \dots j_k} = \langle \Phi | A^{i_1\dots i_k}_{j_1 \dots j_k} | \Phi \rangle$ is a $k$-body density matrix. The braces $\{\cdots\}$ denote normal ordering with respect to the reference state. By construction, the expectation value of any normal-ordered operator in the reference state vanishes,
\begin{equation}
     \langle \Phi | \{ A^{i_1\dots i_k}_{j_1 \dots j_k} \}| \Phi \rangle = 0.
\end{equation}
Consequently, any $k$-body operator $\hat{O}$ can be decomposed exactly into a sum of normal-ordered zero- to $k$-body components:
\begin{equation}
O = {O}^{(0)} +{O}^{(1)}+ \cdots + {O}^{(k)},
\end{equation}
where the normal-ordered $k$-body part is
\begin{equation}
{O}^{(k)} = \frac{1}{(k!)^2} \sum_{i_1\dots i_k, j_1 \dots j_k} {O}^{i_1\dots i_k}_{j_1 \dots j_k}  \{ A^{i_1\dots i_k}_{j_1 \dots j_k} \}.
\end{equation}
We can now normal order the Hamiltonian to obtain
\begin{equation}
    H = E + \sum_{ij}f^{i}_{j}\{A^{i}_{j}\} + \frac{1}{4}\sum_{ijkl}\Gamma^{ij}_{kl}\{A^{ij}_{kl}\} + \frac{1}{36}\sum_{ijklmn}W^{ijk}_{lmn}\{A^{ijk}_{lmn}\}.
\end{equation}
The zero-body part is the expected energy value of the reference state
\begin{equation}
    E = \sum_{ij} t^{i}_{j} \rho^{i}_{j} + \frac{1}{4}\sum_{ijkl} v^{ij}_{kl} \rho^{ij}_{kl} + \frac{1}{36}\sum_{ijklmn} w^{ijk}_{lmn} \rho^{ijk}_{lmn},
\end{equation}
and the one-, two- and three-body terms are respectively
\begin{equation}
    f^{i}_{j} = t^{i}_{j} + \sum_{ab} v^{ia}_{jb} \rho^{a}_{b} + \frac{1}{4}\sum_{abcd} w^{iab}_{jcd} \rho^{ab}_{cd},
\end{equation}

\begin{equation}
    \Gamma^{ij}_{kl} = v^{ij}_{kl}   + \sum_{ab} w^{ija}_{klb} \rho^{a}_{b} ,
\end{equation}

\begin{equation}
    W^{ijk}_{lmn} = w^{ijk}_{lmn} .
\end{equation}

To facilitate the subsequent discussion, we introduce the irreducible $k$-body density matrices $\lambda^{(k)}$. At the one-body level, this is simply the standard density matrix
\begin{equation}
\lambda_{j}^{i} =\rho^i_j.
\end{equation}
It is also useful to define the complementary matrix
\begin{equation}
\xi_{j}^{i} =\lambda_{j}^{i} - \delta_{j}^{i}.
\end{equation}
Up to a sign factor $(-1)$ that unifies the sign conventions for one-body contractions introduced later, $\xi^{(1)}$ generalizes the hole density matrix for a correlated state.  In this work, we are working in the natural orbital basis, in which both $\lambda^{(1)}$ and $\xi$ matrices are diagonal,
\begin{equation}
\label{eq: NAT basis}
\lambda_{j}^{i} = n_{i}\delta_{j}^{i}, \qquad \xi_{j}^{i} = -\bar{n}_{i}\delta_{j}^{i} = -(1-n_{i})\delta_{j}^{i},
\end{equation}
where the eigenvalues $0 \leq n_{i} \leq 1$ are the (potentially fractional) occupation numbers in the reference state $\ket{\Phi}$.

For $k \geq 2$, the irreducible density matrices are defined recursively:
\begin{equation}
\lambda_{kl}^{ij} =\rho_{kl}^{ij} - \mathcal{A} [\lambda_{k}^{i} \lambda_{l}^{j}],
\end{equation}
\begin{equation}
\lambda_{lmn}^{ijk} = \rho_{lmn}^{ijk} - \mathcal{A} [ \lambda_{l}^{i} \lambda_{mn}^{jk} ] - \mathcal{A} [ \lambda_{l}^{i} \lambda_{m}^{j} \lambda_{n}^{k} ],
\end{equation}
and so forth. The irreducible $k$-body density matrix $\lambda^{(k)}$ encodes the genuine $k$-nucleon correlations in the reference state $|\Phi\rangle$. If the reference state has no correlation, e.g., in the case of a Slater determinant (an independent-particle state), the full $k$-body density matrix factorizes completely, and $\lambda^{(k)}$ vanishes for all $k \ge 2$. This defines the SR framework. In contrast, when the reference state is correlated, the irreducible density matrices satisfy $\lambda^{(k\ge 2)} \ne 0$, defining the multi-reference (MR) framework. In the present work, we focus on the more general MR framework; SR expressions can be obtained easily by setting terms with irreducible density matrices to zero.

As we calculate the commutators of normal-ordered many-body operators, we first need to calculate their products, which obey the generalized Wick theorem of Kutzelnigg and Mukherjee \cite{Kutzelnigg1997}. An algebraic proof (and further generalization) can be found in Ref~\cite{Kong2010}. Any product of two normal-ordered operators can be expanded in a sum of normal-ordered terms, with Wick contractions and operators containing at least one index from each of the original operators. For example, the basic contractions appearing in the expansion of a product of normal-ordered two-body operators are (notice the signs):
\begin{subequations}
\label{eq:contraction}
\begin{align}
\{A^{\underline{a}b}_{cd}\}\{A^{ij}_{\underline{k}l}\} 
&= \lambda^{a}_{k} \{ A^{bij}_{cdl} \},\\
\{A_{\underline{c}d}^{ab}\}\{A_{kl}^{i\underline{j}}\} 
&= -\xi^{j}_{c} \{ A^{abi}_{dkl} \},\\
\{A^{\underline{ab}}_{cd}\}\{A^{ij}_{\underline{kl}}\} 
&= +\lambda^{ab}_{kl} \{ A^{ij}_{cd} \},\\
\{A^{\underline{ab}}_{c\underline{d}}\}\{A^{\underline{i}j}_{\underline{kl}}\} 
&= -\lambda^{abi}_{dkl} \{ A^{j}_{c} \}, \\
\{A^{\underline{ab}}_{\underline{cd}}\}\{A^{\underline{ij}}_{\underline{kl}}\} 
&= +\lambda^{abij}_{cdkl}. 
\end{align}
 \end{subequations}

Only the first two contraction types appear in the regular Wick theorem for uncorrelated reference states (SR framework). The additional contractions increase the number of terms when we expand operator products using the generalized Wick theorem (MR framework).

According to the contraction rules of the generalized Wick theorem, the product of a normal-ordered $m$-body operator and a normal-ordered $n$-body operator can be expanded into a sum of normal-ordered operators ranging from zero-body to ($m+n$)-body. However, when computing their commutator, the highest-order ($m+n$)-body normal-ordered terms cancel upon subtraction, so that only terms from zero-body to ($m+n-1$)-body remain. This can be expressed as
\begin{equation}
    [\{A\}^{(m)},\{A\}^{(n)}]
    \equiv \{A\}^{(m)}\{A\}^{(n)}-\{A\}^{(n)}\{A\}^{(m)} = \sum_{k=0}^{m+n-1} \{A\}^{(k)}.
\end{equation}
In the SR limit, the sum is restricted further, since the lowest possible rank that can be achieved with standard contractions will be $k=|m-n|$.

When constructing the IMSRG flow equations, it is necessary to evaluate commutators of normal-ordered many-body operators $dH/ds = [\eta,H]$. Although the initial Hamiltonian contains at most three-body terms, the evaluation of commutators will induce higher-order normal-ordered many-body terms. Therefore, in practical calculations, the operators in the flow equations must be truncated. Here, we consistently truncate all operators at the NO3B level,
\begin{subequations}
\begin{equation}
    H(s) \approx E(s) + f(s)+\Gamma(s)+W(s),
\end{equation}
\begin{equation}
    \eta(s) \approx \eta^{(1)}(s) + \eta^{(2)}(s) + \eta^{(3)}(s).
\end{equation}
\end{subequations}
The flow equation for the Hamiltonian operator reads
\begin{equation}
\label{eq:flow_H}
    \frac{d}{ds}H(s)\approx \frac{d}{ds}E(s) + \frac{d}{ds}f(s) + \frac{d}{ds}\Gamma(s) + \frac{d}{ds}W(s).
\end{equation}

As an illustrative example, we present a step-by-step derivation of the commutator between two normal-ordered two-body operators, contracting to the zero-body component. Let us first consider the contribution of the commutator between two-body operators to the energy derivative,
\begin{equation}
    \frac{d}{ds}E(s) \;\supset\; 
    \frac{1}{16} \sum_{abcdijkl} \eta^{ab}_{cd}\Gamma^{ij}_{kl}
    \Big[\{A^{ab}_{cd}\}, 
    \{A^{ij}_{kl}\}\Big]^{(0)}.
\end{equation}
Here, we retain only the contributions involving the irreducible one-body density matrices for simplicity. 
Applying the generalized Wick theorem, the zero-body component of the commutator that depends on the one-body density matrices is given by
\begin{align}
        [\{A^{ab}_{cd}\},\{A^{ij}_{kl}\}]^{(0)}
        =&\,(1-P_{ab})(1-P_{ij}) \notag\\
        &\times \bigl(\lambda^{a}_{k}\lambda^{b}_{l}\,\xi^{i}_{c}\xi^{j}_{d} 
        - \xi^{a}_{k}\xi^{b}_{l}\,\lambda^{i}_{c}\lambda^{j}_{d} \bigr),
\end{align}
where $P_{ij}$ denotes the permutation operator exchanging the indices $i$ and $j$. The above expression can be simplified in  the natural orbital basis, in which the irreducible one‑body density matrices are diagonal according to Eq.~(\ref{eq: NAT basis}).  Substituting the diagonal expressions, one obtains
\begin{align}
        [\{A^{ab}_{cd}\},\{A^{ij}_{kl}\}]^{(0)}
        =&\,(1-P_{ab})(1-P_{ij}) \notag\\
        &\times \bigl(n_a n_b \bar{n}_c \bar{n}_d - \bar{n}_a \bar{n}_b n_c n_d\bigr)\,
        \delta^{a}_{k}\delta^{b}_{l}\delta^{i}_{c}\delta^{j}_{d}.
\end{align}
Plugging this expression into the energy derivative and exploiting the antisymmetry of the matrix elements, we obtain
\begin{equation}
\begin{aligned}
   & \frac{1}{16} \sum_{abcdijkl}\eta^{ab}_{cd} \Gamma^{ij}_{kl}
    [\{A^{ab}_{cd}\},\{A^{ij}_{kl}\}]^{(0)}\\
    =&\frac{1}{16} \sum_{abcdijkl}\eta^{ab}_{cd} \Gamma^{ij}_{kl}
    \delta^{a}_{k}\delta^{b}_{l}\delta^{i}_{c}\delta^{j}_{d} \notag\\
    &\times (1-P_{ab})(1-P_{ij}) \notag\\
    &\times \bigl(n_a n_b \bar{n}_c \bar{n}_d - \bar{n}_a \bar{n}_b n_c n_d\bigr) \notag\\
    =&\frac{1}{4} \sum_{abcd}\eta^{ab}_{cd} \Gamma^{cd}_{ab}
    \bigl(n_a n_b \bar{n}_c \bar{n}_d - \bar{n}_a \bar{n}_b n_c n_d\bigr).
\end{aligned}
\end{equation}
Here we have used the antisymmetry of the matrix elements,
\begin{equation}
    \Gamma^{ij}_{kl} = -\Gamma^{ji}_{kl} = -\Gamma^{ij}_{lk} = \Gamma^{ji}_{lk},
\end{equation}
which implies that the sign factors introduced by the permutation operators $(1-P_{ab})$ and $(1-P_{ij})$ cancel, so that each permutation sum effectively contributes a factor of $2$.

Finally, the contribution to the zero-body flow equation from the commutator between two-body operators, retaining only the one-body density matrices, is given by
\begin{equation}
    \frac{dE}{ds}(22,\lambda^{(1)}) 
    = \frac{1}{4} \sum_{abcd}\eta^{ab}_{cd} \Gamma^{cd}_{ab}
    \bigl(n_a n_b \bar{n}_c \bar{n}_d - \bar{n}_a \bar{n}_b n_c n_d\bigr).
\end{equation}

This example highlights the tedious and error-prone nature of manual commutator evaluation. One must systematically account for all possible Wick contractions prescribed by the generalized Wick theorem and subsequently exploit the antisymmetry of the matrix elements to simplify the resulting expressions. As the operator rank increases, the number of terms grows rapidly, rendering manual derivations of higher-order commutators practically infeasible. To overcome this challenge, we have developed the \texttt{qcombo} package, which automates the evaluation, simplification, and output of commutators for general many-body operators.

\section{Package Description}
\label{sec:Package_Description}

\begin{figure*}[t]
    \centering
    \includegraphics[width=0.6\linewidth]{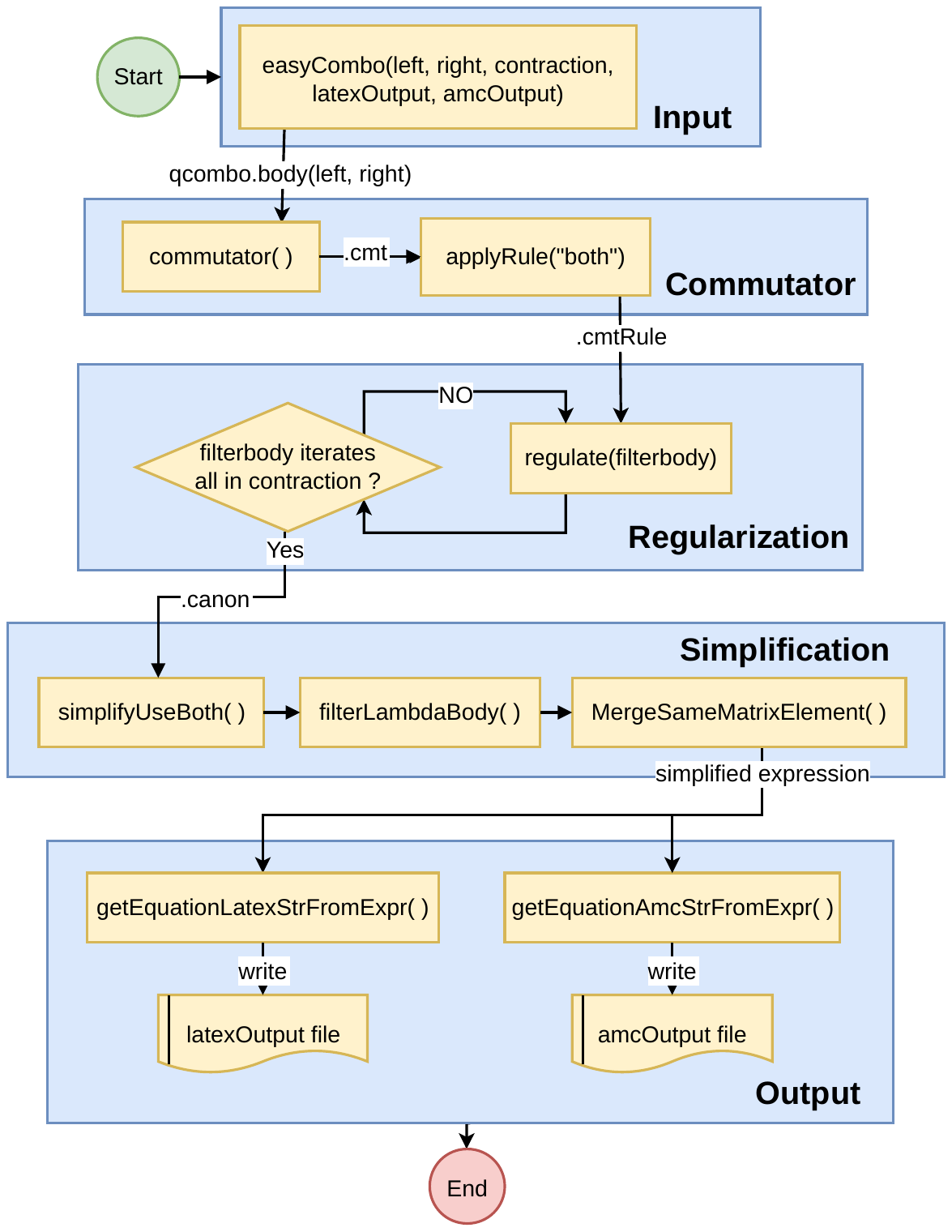}\vspace{-1cm}
    \caption{Flowchart of the main routine {\tt easyCombo} in the {\tt qcombo} package. The workflow includes the construction of commutators, application of contraction rules, regularization of intermediate expressions, simplification of algebraic terms, and generation of outputs in both \LaTeX{} and {\tt AMC} formats. }
    \label{fig:flowchart}
\end{figure*}

In this section, we outline the structure of the {\tt qcombo} package. Its workflow consists of five sequential modules: {\tt input}, {\tt commutator}, {\tt regularization}, {\tt simplification} and  {\tt output}. The {\tt input} module specifies the operators, while the {\tt commutator} module applies the generalized Wick theorem to generate all contractions. The {\tt regularization} module filters terms and enforces a canonical ordering of indices. The {\tt simplification} module further reduces the expressions by exploiting antisymmetry and standardizing the density matrix indices. Finally, the {\tt output} module exports the results in \LaTeX{} format and as input for the {\tt AMC} package. For convenience, these steps are integrated into the {\tt easyCombo} function, which automates the generation of commutator expressions. The workflow is illustrated in Fig.~\ref{fig:flowchart}.

\subsection{Installation}

The {\tt qcombo} package is available on PyPI. It can be installed locally by running the following command in a Python environment with version {\tt Python>=3.12}:
\begin{lstlisting}
pip install qcombo
\end{lstlisting}

Alternatively, the package can be installed directly from the GitHub repository:
\begin{lstlisting}
git clone https://github.com/chenlh73/qcombo.git
cd qcombo
pip install -e .
\end{lstlisting}

  \subsection{Illustration in a Jupyter notebook}

\subsubsection{Input}
First, import the {\tt qcombo} package:
\begin{lstlisting}
import qcombo
\end{lstlisting} 
To display LaTeX expressions in Jupyter notebooks, we define the \lstinline{jupyterDisplay} function. This function first converts \texttt{SymPy} expressions into \LaTeX{} format using \lstinline{texExp}, and then renders them in the notebook through the IPython library. If IPython is not available, the function falls back to directly printing the corresponding \LaTeX{} string.
\begin{lstlisting}
def jupyterDisplay(expr):
    #trans SymPy expression to latex
    latex_expr = qcombo.texExp(expr)
    try:
        from IPython.display import display, Latex
        # display latex in jupyter notebook
        display(Latex(f"$${latex_expr}$$"))
    except ImportError:
        print(latex_expr)
\end{lstlisting}

Next, we define the operator indices involved in the commutator to be evaluated. As an illustrative example, we consider the commutator between a normal-ordered one-body operator and a two-body operator, 
\begin{equation}
  [G^{(1)}, H^{(2)}]
  =\frac{1}{4}\sum_{abijkl} G^a_b H^{ij}_{kl} \Big[\{A^{a}_{b}\}, \{A^{ij}_{kl}\}\Big].
\end{equation} 
The indices associated with the left and right operators must be distinct. Once the indices are specified, the corresponding operator objects can be constructed using the \lstinline{bodys} class for subsequent symbolic manipulations. The \lstinline{bodys} class also provides the following adjustable settings:
\begin{enumerate}
    \item \lstinline{bodys.wick_mode}. Accepts either \lstinline{"SR"} or \lstinline{"MR"} (default). \lstinline{"MR"} indicates a multi‑reference state, allowing contractions involving two‑ and higher-body irreducible density matrices based on the generalized Wick theorem, whereas \lstinline{"SR"} indicates a single‑reference state, allowing contractions involving only one‑body irreducible density matrices.

    \item \lstinline{bodys.show_process}. Default \lstinline{False}; controls whether to display the computation progress.

    \item \lstinline{bodys.parallel}. Default \lstinline{False}; controls whether to enable parallel computation using \lstinline{ProcessPoolExecutor}. Note that for commutators involving operators with low body rank, serial computation may be faster than parallel execution.
\end{enumerate}

 \begin{lstlisting}[mathescape=true,upquote=true]
#[1B,2B]
left=[['a'],['b']]
right=[['i','j'],['k','l']]

#define the bodys
operators = qcombo.bodys(left,right)

# optional settings
operators.wick_mode = "MR"
operators.show_process = True
operators.parallel = False
\end{lstlisting}

\subsubsection{Commutator}

After defining the operators, their commutator is evaluated using the function  \lstinline{Wick.commutator}, which computes the commutator of the two operators via the following steps:
\begin{enumerate}
    \item Call \lstinline{Wick.generalizedWick} to obtain the expression for the product of left and right.
    
    \item  Swap the two operators and call \lstinline{Wick.generalizedWick} again to obtain the product of right and left.
    
    \item Subtract the second expression from the first one, yielding the commutator of left and right.
    
\end{enumerate}
In the above procedure, the \lstinline{Wick.generalizedWick} function computes the product of normal-ordered operators based on the generalized Wick theorem. Its procedure can be summarized as follows:
\begin{enumerate}
\item Identify the upper and lower indices of the two normal-ordered operators. For example, for the product $\{A^a_b\}\{A^{ij}_{kl}\}$, the upper indices are $(a,i,j)$ and the lower indices are $(b,k,l)$. All possible contractions can then be constructed according to the rules in Eq.~(\ref{eq:contraction}). For instance, the index pairing $(a,k)(ij,bl)$ generates terms such as $\lambda^a_k A^{ij}_{bl}$.

\item Determine the sign of each term based on the ordering of the contracted indices, and sum all contributions to obtain the final result.

\end{enumerate}

During the above calculations, the product expression based on the generalized Wick theorem is stored in the {\tt .gw} attribute of the class, while the final commutator expression is stored in {\tt .cmt} attribute. 

\begin{lstlisting}[mathescape=true]
#compute the commutator
operators.commutator()
#the product 
jupyterDisplay(operators.gw)
#the commutator 
jupyterDisplay(operators.cmt)

output:
$A^{a i}_ {k l}\xi^{j}_ {b} + A^{a i j}_ {b k l} - A^{a j}_ {k l}\xi^{i}_ {b} - A^{a}_ {k}\lambda^{i j}_ {b l} + A^{a}_ {l}\lambda^{i j}_ {b k} + A^{i j}_ {b k}\lambda^{a}_ {l} - A^{i j}_ {b l}\lambda^{a}_ {k} - A^{i}_ {b}\lambda^{a j}_ {k l} $
$+ A^{i}_ {k}\lambda^{a j}_ {b l} - A^{i}_ {k}\lambda^{a}_ {l}\xi^{j}_ {b} - A^{i}_ {l}\lambda^{a j}_ {b k} + A^{i}_ {l}\lambda^{a}_ {k}\xi^{j}_ {b} + A^{j}_ {b}\lambda^{a i}_ {k l} - A^{j}_ {k}\lambda^{a i}_ {b l} + A^{j}_ {k}\lambda^{a}_ {l}\xi^{i}_ {b} $
$+ A^{j}_ {l}\lambda^{a i}_ {b k} - A^{j}_ {l}\lambda^{a}_ {k}\xi^{i}_ {b} + \lambda^{a i}_ {k l}\xi^{j}_ {b} + \lambda^{a i j}_ {b k l} - \lambda^{a j}_ {k l}\xi^{i}_ {b} - \lambda^{a}_ {k}\lambda^{i j}_ {b l} + \lambda^{a}_ {l}\lambda^{i j}_ {b k} $

$-A^{a i}_ {k l}\lambda^{j}_ {b} + A^{a i}_ {k l}\xi^{j}_ {b} + A^{a j}_ {k l}\lambda^{i}_ {b} - A^{a j}_ {k l}\xi^{i}_ {b} + A^{i j}_ {b k}\lambda^{a}_ {l} - A^{i j}_ {b k}\xi^{a}_ {l} - A^{i j}_ {b l}\lambda^{a}_ {k} + A^{i j}_ {b l}\xi^{a}_ {k} $
$- A^{i}_ {k}\lambda^{a}_ {l}\xi^{j}_ {b} + A^{i}_ {k}\lambda^{j}_ {b}\xi^{a}_ {l} + A^{i}_ {l}\lambda^{a}_ {k}\xi^{j}_ {b} - A^{i}_ {l}\lambda^{j}_ {b}\xi^{a}_ {k} + A^{j}_ {k}\lambda^{a}_ {l}\xi^{i}_ {b} - A^{j}_ {k}\lambda^{i}_ {b}\xi^{a}_ {l} - A^{j}_ {l}\lambda^{a}_ {k}\xi^{i}_ {b} $
$+ A^{j}_ {l}\lambda^{i}_ {b}\xi^{a}_ {k} - \lambda^{a i}_ {k l}\lambda^{j}_ {b} + \lambda^{a i}_ {k l}\xi^{j}_ {b} + \lambda^{a j}_ {k l}\lambda^{i}_ {b} - \lambda^{a j}_ {k l}\xi^{i}_ {b} - \lambda^{a}_ {k}\lambda^{i j}_ {b l} + \lambda^{a}_ {l}\lambda^{i j}_ {b k} - \lambda^{i j}_ {b k}\xi^{a}_ {l} $
$+ \lambda^{i j}_ {b l}\xi^{a}_ {k}$
\end{lstlisting}
Note that the symbol $A^{ab\cdots}_{ij\cdots}$ in the output denotes the normal-ordered operator $\{A^{ab\cdots}_{ij\cdots}\}$. All indices are understood to be contracted either among themselves or with those of the matrix elements of the two operators.

After obtaining the commutator expression, we invoke the function \lstinline{applyRule(ruleType)} to simplify it. This function internally processes the {\tt .cmt} expression from the previous step, with its behavior controlled by the \lstinline{ruleType} parameter, which supports three different simplification modes:  'xi', 'nat', and 'both'. The rules for each mode are as follows:
\begin{enumerate}
    \item \lstinline{ruleType='xi'}. This mode replaces all occurrences of the $\xi$ operator according to its definition: $\xi^i_j = \lambda^i_j - \delta^i_j$.

    \item \lstinline{ruleType='nat'}. In this mode, the one‑body irreducible density matrix is diagonalized on the natural‑orbital basis using the relation $\lambda^i_j = n_i \delta^i_j$.

    \item \lstinline{ruleType='both'} (default).
   This mode first applies the 'xi' substitution and then performs the 'nat' diagonalization, effectively combining both rules described above.
\end{enumerate}

After simplification, the irreducible one‑body density matrices are eliminated and replaced by the occupation numbers $n_i$ and Kronecker‑$\delta$ symbols. The resulting expression is stored in the {\tt .cmtRule} attribute.

\begin{lstlisting}[mathescape=true]
#diagonalize the xi and lambda
operators.applyRule(ruleType='both')
#the diagonalized commutator
jupyterDisplay(operators.cmtRule)

output:
$-A^{a i}_ {k l}\delta^{j}_ {b} + A^{a j}_ {k l}\delta^{i}_ {b} + A^{i j}_ {b k}\delta^{a}_ {l} - A^{i j}_ {b l}\delta^{a}_ {k} + A^{i}_ {k}n^{}_ {a}\delta^{a}_ {l}\delta^{j}_ {b} - A^{i}_ {k}n^{}_ {j}\delta^{a}_ {l}\delta^{j}_ {b} - A^{i}_ {l}n^{}_ {a}\delta^{a}_ {k}\delta^{j}_ {b} $
$+ A^{i}_ {l}n^{}_ {j}\delta^{a}_ {k}\delta^{j}_ {b} - A^{j}_ {k}n^{}_ {a}\delta^{a}_ {l}\delta^{i}_ {b} + A^{j}_ {k}n^{}_ {i}\delta^{a}_ {l}\delta^{i}_ {b} + A^{j}_ {l}n^{}_ {a}\delta^{a}_ {k}\delta^{i}_ {b} - A^{j}_ {l}n^{}_ {i}\delta^{a}_ {k}\delta^{i}_ {b} - \delta^{a}_ {k}\lambda^{i j}_ {b l} $
$+ \delta^{a}_ {l}\lambda^{i j}_ {b k} + \delta^{i}_ {b}\lambda^{a j}_ {k l} - \delta^{j}_ {b}\lambda^{a i}_ {k l}$

\end{lstlisting}

\subsubsection{Regularization}

After obtaining the expression for the commutator of normal-ordered operators, it is necessary to multiply it by the corresponding matrix elements. This step is implemented by the \lstinline{regulate(filterbody)} function, which performs three main tasks: multiplying the previously derived {\tt .cmtRule} commutator expression by the matrix elements, filtering the terms according to the desired many-body rank, and finally simplifying the expression by contracting Kronecker delta indices. The detailed procedure is as follows:
\begin{enumerate}
    \item  Multiplication by matrix elements of operators.  
    The {\tt .cmtRule} expression is passed internally and multiplied by the interaction matrix elements. Note that  the prefactors of all operators are dropped at this stage and they will be included later. In the code, the matrix element for the left operator is denoted by "G", and for the right operator by "H". The indices of these matrix elements correspond to those initially provided when defining the operators.
    
    \item Filtering by the properties of operators. Specific contraction terms are then filtered based on the number of indices attached to the operator symbol "A". This filtering is controlled by the \lstinline{filterbody} parameter. By default (\lstinline{filterbody=None}), all many-body terms are retained. When \lstinline{filterbody=k} (where k is an integer) is specified, only expressions that contract to the $k$-body operator are output and stored in the {\tt .filterTerms} attribute.
    
    \item Simplification and re-indexing. The program simplifies the filtered expression by contracting indices via the Kronecker deltas. New dummy summation indices are then generated, starting from "a" and proceeding in alphabetical order. These indices are preferentially assigned to the normal-ordered operator symbols "A". This ensures that within the same many-body term, the indices of the normal-ordered operators remain consistent. The final, canonicalized expression is stored in the {\tt .canon} attribute.
    
\end{enumerate}
Consider the commutator of one-body and two-body operators derived previously, which yields contractions from zero-body to two-body terms. By setting the parameter \lstinline{filterbody=1}, we can filter out and obtain the expression for all the one-body terms:

\begin{lstlisting}[mathescape=true]
#filter [1B,2B]-1B and normalize the indices
operators.regulate(filterbody=1)
#filtered terms with Matrix elements
jupyterDisplay(operators.filterTerms)
#contract the delta and canonicalize indices
jupyterDisplay(operators.canon)

output:
$A^{i}_ {k}G^{a}_ {b}H^{i j}_ {k l}n^{}_ {a}\delta^{a}_ {l}\delta^{j}_ {b} - A^{i}_ {k}G^{a}_ {b}H^{i j}_ {k l}n^{}_ {j}\delta^{a}_ {l}\delta^{j}_ {b} - A^{i}_ {l}G^{a}_ {b}H^{i j}_ {k l}n^{}_ {a}\delta^{a}_ {k}\delta^{j}_ {b} + A^{i}_ {l}G^{a}_ {b}H^{i j}_ {k l}n^{}_ {j}\delta^{a}_ {k}\delta^{j}_ {b} $
$- A^{j}_ {k}G^{a}_ {b}H^{i j}_ {k l}n^{}_ {a}\delta^{a}_ {l}\delta^{i}_ {b} + A^{j}_ {k}G^{a}_ {b}H^{i j}_ {k l}n^{}_ {i}\delta^{a}_ {l}\delta^{i}_ {b} + A^{j}_ {l}G^{a}_ {b}H^{i j}_ {k l}n^{}_ {a}\delta^{a}_ {k}\delta^{i}_ {b} - A^{j}_ {l}G^{a}_ {b}H^{i j}_ {k l}n^{}_ {i}\delta^{a}_ {k}\delta^{i}_ {b}$

$A^{a}_ {b}G^{c}_ {d}H^{a d}_ {b c}n^{}_ {c} - A^{a}_ {b}G^{c}_ {d}H^{a d}_ {b c}n^{}_ {d} - A^{a}_ {b}G^{c}_ {d}H^{a d}_ {c b}n^{}_ {c} + A^{a}_ {b}G^{c}_ {d}H^{a d}_ {c b}n^{}_ {d} - A^{a}_ {b}G^{c}_ {d}H^{d a}_ {b c}n^{}_ {c} $
$+ A^{a}_ {b}G^{c}_ {d}H^{d a}_ {b c}n^{}_ {d} + A^{a}_ {b}G^{c}_ {d}H^{d a}_ {c b}n^{}_ {c} - A^{a}_ {b}G^{c}_ {d}H^{d a}_ {c b}n^{}_ {d}$
\end{lstlisting}

\subsubsection{Simplification}

After filtering and canonical re-indexing, the number of terms in the expression grows exponentially with the rank of the many-body operators, necessitating further simplification. The simplification module addresses this by leveraging the antisymmetry of matrix element indices and the renaming property of dummy summation indices to reduce the total number of terms, combine like terms, and produce a more concise final expression.

The first step of simplification is implemented through three core functions: \lstinline{simplifyUseAntisymmetry(expr)}, \lstinline{simplifyUseDummyIndices(expr)}, and \lstinline{simplifyUseBoth(expr)}, where \lstinline{expr} is the input expression to be simplified. Their underlying principles are as follows:
\begin{enumerate}
    \item \lstinline{simplifyUseAntisymmetry(expr)}. This function utilizes the antisymmetry property of the matrix element indices. For each term in the expression, the indices of two-body (and higher) operators are rearranged into a canonical order based on alphabetical sorting. A corresponding sign factor, determined by the parity of the index permutations, is applied. The primary benefit is that matrix elements which are identical up to a permutation of their upper and lower indices will obtain a consistent canonical ordering, allowing like terms to be identified and combined. For example, 
    \begin{align}
        G^{ia}_{kc}H^{jb}_{ld} &= (-1)^4 G^{ai}_{ck}H^{bj}_{dl} \\
        -G^{ai}_{kc}H^{jb}_{ld} &= -(-1)^3 G^{ai}_{ck}H^{bj}_{dl}.
    \end{align}
    The terms $G^{ia}_{kc}H^{jb}_{ld}$ and $-G^{ai}_{kc}H^{jb}_{ld}$ appear different and would not be combined initially. By applying antisymmetric reordering, we find they are identical and can be merged to reduce the total term count.

    \item \lstinline{simplifyUseDummyIndices(expr)}. This function exploits the renamability of dummy summation indices. It performs a self-mapping on the set of dummy indices, preferentially assigning them to irreducible density matrices of rank two and higher. This ensures a consistent index labeling for density matrices, facilitating further simplification and yielding a cleaner expression. As an example, consider
    \begin{align}
    A^a_b G^{cd}_{ef} H^{ag}_{bc} \lambda^{dg}_{ef} &\to A^a_b G^{gc}_{ef} H^{ad}_{bg} \lambda^{cd}_{ef} \quad (\text{map } dgefc \to cdefg)\,, \\
    A^a_b G^{gd}_{ef} H^{ac}_{bg} \lambda^{dc}_{ef} &\to A^a_b G^{gc}_{ef} H^{ad}_{bg} \lambda^{cd}_{ef} \quad (\text{map } dcefg \to cdefg)\,.
    \end{align}
    Although the terms $A^a_b G^{cd}_{ef} H^{ag}_{bc} \lambda^{dg}_{ef}$ and $A^a_b G^{gd}_{ef} H^{ac}_{bg} \lambda^{dc}_{ef}$ appear distinct, they become identical upon relabeling the summation indices. Therefore, the two terms are equivalent and can be combined.

  \item \lstinline{simplifyUseBoth(expr)}. This function combines the two approaches above. It first applies \lstinline{simplifyUseAntisymmetry} to reduce the number of terms, followed by \lstinline{simplifyUseDummyIndices} to identify equivalent contributions via index relabeling. Since this step may disrupt the canonical ordering of indices, \lstinline{simplifyUseAntisymmetry} is applied once more to restore a canonical form.
     
\end{enumerate}
Let us demonstrate the application of \lstinline{simplifyUseBoth} to simplify the previous commutator expression: 

\begin{lstlisting}[mathescape=true]
#unsimplified expression
initial_expr = operators.canon
#simplify the expression
simplified_expr = qcombo.simplifyUseBoth(initial_expr)
jupyterDisplay(simplified_expr)

output:
$4A^{a}_ {b}G^{c}_ {d}H^{a d}_ {b c}n^{}_ {c} - 4A^{a}_ {b}G^{c}_ {d}H^{a d}_ {b c}n^{}_ {d}$
\end{lstlisting}
 
When commutator expressions involve higher-body operators, they generally contain irreducible density matrices of corresponding rank. To keep the expressions concise and manageable, it is useful to filter terms according to the body of the irreducible density matrices they contain. This functionality is provided by the \lstinline{filterLambdaBody(expr, filterLambdaBody)} function, where \lstinline{expr} denotes the input expression and \lstinline{filterLambdaBody} specifies the rank of the irreducible density matrices we want to retain.

The function scans each term and selects those containing irreducible density matrices of the specified rank, determined by the number of indices on the $\lambda$ symbol. All other terms are discarded.

We apply this filter to the simplified expression, retaining only terms involving the irreducible one-body density matrix. Since the original expression already contains only one-body density matrices, the filtered result coincides with the input.

\begin{lstlisting}[mathescape=true]
#filter 1B-lambda terms
filter_lambda_expr = qcombo.filterLambdaBody(simplified_expr,LambdaBody=1)
jupyterDisplay(filter_lambda_expr)

output:
$4A^{a}_ {b}G^{c}_ {d}H^{a d}_ {b c}n^{}_ {c} - 4A^{a}_ {b}G^{c}_ {d}H^{a d}_ {b c}n^{}_ {d}$
\end{lstlisting}

After simplification and filtering, expressions may still contain terms in which the matrix elements of the left and right operators share identical indices. In such cases, the built-in \lstinline{simplify} function in {\tt SymPy} is unable to combine these terms effectively. To address this issue, {\tt qcombo} provides the \lstinline{MergeSameMatrixElement(expr)} function, where \lstinline{expr} denotes the input expression.
This function iterates over all terms and identifies those containing matrix elements (denoted by "G" and "H") with identical index structures. The corresponding coefficients are extracted and summed, yielding a single combined coefficient. The final result is a simplified expression in which like terms are properly merged.

We apply this function to the simplified expression of the one-body contraction involving the one-body irreducible density matrix. The resulting expression is more compact, with equivalent terms consistently combined.

\begin{lstlisting}[mathescape=true]
#merge the same matrix element
united_expr = qcombo.MergeSameMatrixElement(filter_lambda_expr)
jupyterDisplay(united_expr)

output:
$4(n^{}_ {c} - n^{}_ {d})A^{a}_ {b}G^{c}_ {d}H^{a d}_ {b c}$
\end{lstlisting}

\subsubsection{Output}

For an arbitrary many-body commutator $ R^{(k)} = [G^{(m)}, H^{(n)}]^{(k)} $ that contracts to a k-body operator, we aim to obtain the expression for the matrix element of the resulting k-body operator. In the code, this is achieved by the function \lstinline{getEquationLatexStrFromExpr(expr, left_body, right_body, contractionBase)}. Here, \lstinline{expr} is the input expression for the $k$-body contracted term; \lstinline{left_body} and \lstinline{right_body} characterize the left and right operators in the commutator, respectively; and \lstinline{contractionBase} is the symbol used for the matrix element of the resulting contracted operator. The function converts the contracted expression into an explicit equation for the matrix element. Its workflow is as follows:
\begin{enumerate}
    \item From the indices of the operator symbol “A” in the expression, identify the output indices of the contracted term; the remaining indices are treated as summation indices.
    
    \item Note that the combinatorial prefactors arising from the definitions of the operators were not taken into account in the preceding derivation. These factors must be included in the final expression. In general, for a $k$-body operator induced from the commutator of  an $m$-body and an $n$-body operator, an additional multiplicative factor of  $\dfrac{(k!)^2}{(m!)^2 (n!)^2}$ is required.  As a specific example,  for $m=1, n=2$ and $k=1$, this   factor is 1/4, which exactly cancels the factor of $4$. 

    \item Both sides of the equation are converted to strings and returned.
\end{enumerate}
As an illustration, we assign the output matrix element for the commutator $[G^{(1)}, H^{(2)}]^{(1)}$ to the symbol ``R'':

\begin{lstlisting}[mathescape=true]
from qcombo.output import getEquationLatexStrFromExpr
#convert expresion to latex equation 
latex_str = getEquationLatexStrFromExpr(united_expr,1,2,'R')
print(latex_str)

output:
R^{a}_{b} = \sum_{c d} (n^{}_{c}-n^{}_{d}) \* G^{c}_{d} \* H^{ad}_{bc}
\end{lstlisting}
Thus, one obtains the one-body operator induced from the commutator of $G^{(1)}$ and $H^{(2)}$ as
\begin{equation}
[G^{(1)}, H^{(2)}]^{(1)}
= \sum_{ab} R^a_b\{A^{a}_{b}\},
\end{equation}
with matrix elements given by
\begin{equation}
R^a_b
= \sum_{cd}(n_c - n_d) G^{c}_{d} H^{ad}_{bc}.
\end{equation}
If the operator $G$ is anti-Hermitian and 
$H$ is Hermitian, one can readily verify from the above expression that $R^a_b=R^b_a$. 

The procedure described above yields commutator expressions in the $M$‑scheme. In practical applications, however, one often needs to work in the angular‑momentum coupled scheme($J$‑scheme). This conversion can be performed automatically using the {\tt AMC} package, which takes an M‑scheme expression as input and outputs the corresponding coupled $J$‑scheme expression. More details about the usage of {\tt AMC} package can be found in Ref.~\cite{Tichai2020amc}.

To interface with {\tt AMC}, we provide the function \lstinline{getEquationAmcStrFromExpr(expr, left_body, right_body, contractionBase_str)}. This function converts an input expression into an equation string that conforms to the {\tt AMC} input format. Its parameters and internal workflow are identical to those of \lstinline{getEquationLatexStrFromExpr}, except that the final string is formatted according to the {\tt AMC} syntax instead of \LaTeX{}.
\begin{lstlisting}
from qcombo.output import getEquationAmcStrFromExpr
#trans expresion to amc equation
amc_str = getEquationAmcStrFromExpr(united_expr,1,2,'R')
print(amc_str)

output:
R1_ab = 1/4*sum_cd(4*(n_c-n_d)*G_cd*H_adbc);
\end{lstlisting}
A notable difference from the \LaTeX format output is that the {\tt AMC} output automatically generates the symbol ``R1'' instead of ``R''. The appended ``1'' indicates that this is a one‑body operator, a convention used when declaring operators in {\tt AMC} input files.

Note that the above procedure assumes all operators are scalar. The situation becomes slightly more involved when tensor operators are considered~\cite{Zhou:2026}. Nevertheless, this case can still be treated using {\tt qcombo} and {\tt AMC}.

\subsection{One-click execution}

To facilitate ease of use, we have integrated the steps described above into a unified function:  

\begin{lstlisting}
easyCombo(left, right, contraction=None, latexOutput=None, amcOutput=None, **kwargs) 
\end{lstlisting}

This function takes as input the body orders (or explicit indices) of the left and right operators, and executes the entire workflow discussed in the preceding sections from start to finish. This includes evaluating the corresponding commutator, generating the required contraction terms, and writing the resulting expressions, both in \LaTeX{} and {\tt AMC} formats, to the specified output files. 

The input parameters are defined as follows:
\begin{enumerate}
    \item \lstinline{left}. The body of the left operator in the commutator.  Accepts either an integer or a list of strings representing the operator indices, e.g., \lstinline{1} or \lstinline{[['a'],['b']]} both denote a normal-ordered one‑body operator.

    \item \lstinline{right}. The body of the right operator in the commutator, with the same format as parameter \lstinline{left}.

    \item \lstinline{contraction}. The desired contraction terms to output. Accepts an integer or a list of integers, e.g., \lstinline{0} or \lstinline{[0, 1, ...]}. The default \lstinline{None} outputs all possible contraction terms.

    \item \lstinline{latexOutput}. Filename for the \LaTeX{}‑formatted equation output. When set to \lstinline{None} (default), the corresponding filename is automatically generated based on the \lstinline{left}, \lstinline{right}, and \lstinline{contraction} parameters.

    \item \lstinline{amcOutput}. Filename for the {\tt AMC}‑formatted equation output. When set to \lstinline{None} (default), the corresponding filename is automatically generated based on the \lstinline{left}, \lstinline{right}, and \lstinline{contraction} parameters.
    
\end{enumerate}

The optional keyword arguments are as follows:
\begin{enumerate}
\item \lstinline{wick_mode}. Specifies the reference-state mode for Wick contractions. The options are either \lstinline{"SR"} or \lstinline{"MR"} (default). The \lstinline{"SR"} mode denotes a single-reference state, which restricts contractions to one-body irreducible density matrices. The \lstinline{"MR"} mode denotes a multi-reference state, which permits contractions involving many-body irreducible density matrices.

\item \lstinline{show_process}. The default is \lstinline{True}. This flag determines whether the computation progress is displayed.

\item \lstinline{parallel}. The default is \lstinline{False}. This flag determines whether parallel computation is enabled.

\item \lstinline{savefile}. The default is \lstinline{True}. This flag determines whether the \LaTeX{} output (as a \texttt{.tex} file) and the \texttt{AMC} output (as a \texttt{.amc} file) are saved.
\end{enumerate}

As an illustrative example, we consider the contraction of two two-body operators yielding a zero-body term.
\begin{lstlisting}
# [2B,2B]-0B
qcombo.easyCombo(2,2,0)
\end{lstlisting}

Executing the corresponding code yields two files with default names: \texttt{commutator\_2B2B\_to\_0B.tex} and \texttt{commutator\_2B2B\_to\_0B.amc}. After compilation, the contents of the \LaTeX{} file \texttt{commutator\_2B2B\_to\_0B.tex} are displayed as follows:
\begin{itemize}
\item[(1)] one-body irreducible density matrix $\lambda^{(1)}$
\begin{eqnarray}
    R^{}_{} &=& \sum_{abcd} (-n^{}_{a} \* n^{}_{b} \* n^{}_{c}-n^{}_{a} \* n^{}_{b} \* n^{}_{d}+n^{}_{a} \* n^{}_{b}+n^{}_{a} \* n^{}_{c} \* n^{}_{d} \nonumber\\
    &&+n^{}_{b} \* n^{}_{c} \* n^{}_{d}-n^{}_{c} \* n^{}_{d}) \* G^{ab}_{cd} \* H^{cd}_{ab}/4
\end{eqnarray}

\item[(2)] two-body irreducible density matrix $\lambda^{(2)}$ 
\begin{align*}
     R^{}_{} &= \sum_{abcdef} \Bigg[8 \* (n^{}_{e}-n^{}_{f}) \* G^{ae}_{cf} \* H^{bf}_{de}-(n^{}_{e}+n^{}_{f}-1) \* G^{ab}_{ef} \* H^{ef}_{cd} \\
    &+(n^{}_{e}+n^{}_{f}-1) \* G^{ef}_{cd} \* H^{ab}_{ef} \Bigg] \lambda^{ab}_{cd}/8 
\end{align*} 

\item[(3)] three-body irreducible density matrix $\lambda^{(3)}$
\begin{equation*}
    R^{}_{} = \sum_{abcdefg} (G^{ag}_{de} H^{bc}_{fg} - G^{ab}_{dg}  H^{cg}_{ef}) \lambda^{abc}_{def}/4
\end{equation*}
    
\end{itemize}

The generated {\tt AMC} file reads as follows:
\begin{lstlisting}
declare G{mode= (2,2),latex ="G" } 
declare H{mode= (2,2),latex ="H" } 
declare R0{mode= (0,0),latex ="R" } 
declare lambda2B{mode= (2,2),latex ="\lambda" } 
declare lambda3B{mode= (3,3),latex ="\lambda" } 
declare n {  mode=2, diagonal=true, latex="n"} 
 
# commutator [2B,2B]-0B 
# lambda_1B
R0 = 1/16*sum_abcd(4*(-n_a*n_b*n_c-n_a*n_b*n_d+n_a*n_b+n_a*n_c*n_d+n_b*n_c*n_d-n_c*n_d)*G_abcd*H_cdab);
 
# lambda_2B
R0 = 1/16*sum_abcdef(2*(8*(n_e-n_f)*G_aecf*H_bfde-(n_e+n_f-1)*G_abef*H_efcd+(n_e+n_f-1)*G_efcd*H_abef)*lambda2B_abcd);
 
# lambda_3B
R0 = 1/16*sum_abcdefg(4*(-G_abdg*H_cgef+G_agde*H_bcfg)*lambda3B_abcdef);
\end{lstlisting}

Note that some coefficients in the expressions can become rather lengthy. These can be further simplified by exploiting algebraic relations among the occupation numbers, which make the underlying index symmetries more transparent. In particular, one can use the identities
\begin{align}
    &(-n_a n_b n_c - n_a n_b n_d + n_a n_b + n_a n_c n_d + n_b n_c n_d - n_c n_d) \nonumber \\
    &= n_a n_b \bar{n}_c \bar{n}_d - \bar{n}_a \bar{n}_b n_c n_d
\end{align}
and
\begin{equation}
    (n_e + n_f - 1) = n_e n_f - \bar{n}_e \bar{n}_f,
\end{equation} 
where $\bar{n}_i = 1 - n_i$.

\subsection{Execution in bash environment}
For direct command-line execution, users can compute commutators and generate corresponding output files using the command:
\begin{lstlisting}
qcombo LEFT RIGHT [OPTIONS]
\end{lstlisting}
This command essentially invokes the \lstinline{easyCombo} function with the following arguments:

\begin{enumerate}
    \item \verb|LEFT|. The body of the left operator (as an integer) or its specific index list (provided as a Python list string).

    \item \verb|RIGHT|. The body of the right operator (as an integer) or its specific index list (provided as a Python list string).
\end{enumerate}

\noindent The following optional arguments can be invoked:
\begin{enumerate}
    \item \verb|--contraction CONTRACTION, -c CONTRACTION| Specify the desired contraction body numbers. This can be a single integer (e.g., 0), a range (e.g., 0-2), a list (e.g., 0,1,2), or 'all' to calculate all possible contractions. Default is 'all'.

    \item \verb|--latex-output LATEX_OUTPUT, -lo LATEX_OUTPUT| Specify the filename for the \LaTeX{} output. Defaults to an auto-generated name based on the input operators.

    \item \verb|--amc-output AMC_OUTPUT, -ao AMC_OUTPUT| Specify the filename for the {\tt AMC} program input file. Defaults to an auto-generated name based on the input operators.

    \item \verb|--wick-mode {SR, MR}, -w {SR, MR} | The reference state for Wick contraction. "SR" for single-reference state or "MR" for multi-reference state. Default is "MR".

    \item \verb|--parallel, -p| Enable parallel computation.

    \item \verb|--no-process, -ns| Disable progress output during computation.

    \item \verb|--version, -v| Display the program's version number and exit.

    \item \verb|--interactive -i| Launch interactive mode. Note that running the command without arguments also enters this mode.

    \item  \verb|--help -h| Show the help message and exit.
\end{enumerate}

Based on these options, executing the following command in the terminal
\begin{lstlisting}
qcombo 2 2 -c 0 -lo result.tex -ao result.amc
\end{lstlisting}
produces results consistent with those obtained from the previous call to \lstinline{easyCombo(2, 2, 0)}.

\section{Application to MR-IMSRG(3)}
\label{sec:Application}

In the MR-IMSRG(3), the commutators $[A,B]\to C$  are truncated at the NO3B level. The general expressions for the commutator of $m$- and $n$-body operators, automatically generated and simplified by the {\tt qcombo} package, are provided in ~\ref{app:commutators_NO3B} and labeled as ``mnX'' commutators. For clarity, the coefficients involving one-body operators are rewritten in forms that explicitly exhibit their symmetry properties. In addition, terms in the commutator that contain two two-body irreducible density matrices have been reformulated by exploiting their additional symmetries, making these properties more transparent. Consequently, the expressions presented here differ slightly from the direct output of the program, although they are algebraically equivalent.

In the MR-IMSRG(3) framework, the {\tt qcombo} package generates expressions involving irreducible density matrices up to the five-body level. Specifically, the flow equation for the normal-ordered zero-body term contains irreducible density matrices up to the five-body level. The one-body flow equation includes contributions involving $\lambda^{(3)}$, $\lambda^{(4)}$, and products of the form $\lambda^{(2)}\lambda^{(2)}$, while the two-body flow equation contains irreducible density matrices up to $\lambda^{(3)}$. The three-body flow equation involves irreducible density matrices up to $\lambda^{(2)}$. These represent the highest density ranks allowed by the corresponding NO3B commutators. 
To keep the expressions manageable, the flow equations are decomposed according to the origin of the commutator contributions. Specifically, $df^a_b(mn)$ denotes the contribution arising from the commutator between an $m$-body and an $n$-body operator. Similarly, $d\Gamma^{ab}_{cd}(mn, \lambda^{(k)})$ denotes the component of the same commutator that depends only on the $k$-body irreducible density matrix. We note that, contributions from the higher-body ($k>2$) irreducible density matrices play a potentially important role in the convergence and stability of the flow equations.

The zero-, one-, two-, and three-body flow equations in $H(s)$ [\ref{eq:flow_H}] are summarized
below. 

The zero-body flow equation is

\begin{equation}
\begin{aligned}
\label{eq:IMSRG3-0b}
    \frac{dE}{ds} 
    &= \sum_{ab}(n_a - n_b)\eta^{a}_{b}f^{b}_{a} \\
    &+ \frac{1}{4}\sum_{abcd}(n_a n_b \bar{n}_c \bar{n}_d - \bar{n}_a \bar{n}_b n_c n_d)\eta^{a b}_{c d}\Gamma^{c d}_{a b} \\
    &+ \frac{1}{36} \sum_{abcdef}(n_a n_b n_c \bar{n}_d \bar{n}_e \bar{n}_f - \bar{n}_a \bar{n}_b \bar{n}_c n_d n_e n_f  )\eta^{abc}_{def}W^{def}_{abc} \\
    &+ \frac{1}{4}\sum_{abcd}[d\Gamma^{ab}_{cd}(12) + d\Gamma^{ab}_{cd}(22) \\ 
    &+   d\Gamma^{ab}_{cd}(23,\lambda^{(1)}) +  d\Gamma^{ab}_{cd}(33,\lambda^{(1)})]\lambda^{ab}_{cd} \\ 
    &+ \frac{1}{8}\sum_{abcd}[d\Gamma^{ab}_{cd}(23,\lambda^{(2)})-d\Gamma^{ab}_{cd}
    (23,\lambda^{(2)},{\rm non}) \\
    &+  d\Gamma^{ab}_{cd}(33,\lambda^{(2)}) ]\lambda^{ab}_{cd} \\ 
    &+  \frac{1} {36}\sum_{abcdef}\frac{dW^{abc}_{def}}{ds}\lambda^{abc}_{def} \\
    &+dE(23,\lambda^{(4)}) + dE(33,\lambda^{(4)}) +dE(33,\lambda^{(5)}).
\end{aligned}
\end{equation} 
The subtraction term
$d\Gamma^{ab}_{cd}(23,\lambda^{(2)},{\rm non})$ removes the part that
would otherwise be overcounted when the zero-body $\lambda^{(2)}\lambda^{(2)}$ contribution is reconstructed from the two-body flow. See Eq.(\ref{eq:dGamma23-non}) for details. 

The one-body flow equation is
\begin{equation}
\label{eq:IMSRG3-1b}
\begin{aligned}
    \frac{df^a_b}{ds} &= df^{a}_{b}(11) + df^{a}_{b}(12) + df^{a}_{b}(22) \\
    &+ df^{a}_{b}(13) + df^{a}_{b}(23) + df^{a}_{b}(33). 
\end{aligned}
\end{equation}  
The two-body flow equation is
\begin{equation}
\label{eq:IMSRG3-2b}
        \frac{d\Gamma^{ab}_{cd}}{ds} = d\Gamma^{ab}_{cd}(12)+ d\Gamma^{ab}_{cd}(13) + d\Gamma^{ab}_{cd}(22)  + d\Gamma^{ab}_{cd}(23) + d\Gamma^{ab}_{cd}(33). 
\end{equation}
Finally, the three-body flow equation is
\begin{equation}
\label{eq:IMSRG3-3b}
     \frac{dW^{abc}_{def}}{ds} = dW^{abc}_{def}(13) + dW^{abc}_{def}(22) + dW^{abc}_{def}(23) + dW^{abc}_{def}(33)  .
\end{equation}
Explicit expressions for all contributions are provided in
\ref{app:IMSRG3}.

The above flow equations are benchmarked against the SR-IMSRG(3) results~\cite{Hergert:2016PR,Heinz:2021}, in which all two-body and higher irreducible density matrices vanish. Furthermore, when all three-body operators are neglected, the equations reduce to the MR-IMSRG(2) expressions reported in Ref.~\cite{Hergert:2013PRL}.

The computational time exhibits a steep increase with the rank of the many-body operators. For a workstation equipped with an Intel(R) Core(TM) i5-10200H CPU (2.4 GHz) and an NVIDIA GeForce GTX 1650 GPU, the evaluation times of the 11X, 22X, and 33X commutators are approximately 0.07 s, 2.45 s, and 943 s, respectively, using eight parallel processes. These results illustrate the rapidly increasing complexity associated with higher-rank many-body operators.

\section{Conclusion and outlook}
\label{sec:Conclusion}

We have developed the Python package {\tt qcombo} for the symbolic evaluation of commutators between general many-body operators in normal-ordered form, based on the generalized Wick theorem. The structure and key functionalities of the package have been presented. As a representative application, we have derived the complete set of expressions for the multi-reference in-medium similarity renormalization group (MR-IMSRG) method with operators truncated at the normal-ordered three-body (NO3B) level. The package provides an efficient and systematic framework for handling many-body operator contributions in quantum many-body methods. The resulting expressions are delivered in symbolic form, thereby facilitating subsequent analytical developments and numerical implementations.

In the future, we plan to extend the capabilities of {\tt qcombo} in several directions. First, the current implementation supports commutators involving particle-number-conserving operators, and a natural next step is to incorporate analogous functionality for particle-number-breaking operators. In addition, symbolic-algebra frameworks developed in quantum chemistry, such as the Tensor Contraction Engine (TCE), SeQuant, and SymGen, provide a broader range of capabilities, extending from symbolic many-body algebra to automated generation of scalable numerical code (see Appendix C). Inspired by these developments, we envision extending \texttt{Qcombo} along similar lines while tailoring its future development to the specific requirements of nuclear many-body theory.

\section*{Acknowledgments}
 This work is supported in part by the National Natural Science Foundation of China (Grant Nos. 125B2108 and 12375119). H.H. acknowledges the support of the U.S. Department of Energy, Office of Science, Office of Nuclear Physics under Award Numbers DESC0023516 and DE-SC0023175 (SciDAC-5 NUCLEI Collaboration).

\appendix

\section{Commutators under the NO3B approximation}
\label{app:commutators_NO3B}

In this appendix, we collect the expressions for the commutators of many-body operators within the normal-ordered three-body (NO3B) approximation.

In general, the commutator of two many-body operators, $C = [A,B]$, can be expressed as
\begin{equation}
   O = O_0 + \sum_{ab} O^a_b \{A^a_b\}
   + \frac{1}{4}\sum_{abcd} O^{ab}_{cd} \{A^{ab}_{cd}\}
   + \frac{1}{36}\sum_{abcdef} O^{abc}_{def} \{A^{abc}_{def}\},
\end{equation}
where $O$ denotes a generic operator, representing $A$, $B$, or $C$.

The explicit expressions for all matrix elements of the commutator $C$ are given below.

\begin{itemize} 

\item \textbf{The 11X commutator}

\begin{align}
    C_0 &= \sum_{ab} (n_a - n_b)\, A^{a}_{b} B^{b}_{a}
\end{align}

\begin{align}
    C^a_b &= \sum_{c} \left( A^{a}_{c} B^{c}_{b} - A^{c}_{b} B^{a}_{c} \right)
\end{align}

\item \textbf{The 12X commutator}

\begin{align}
    C_0 &= \frac{1}{2} \sum_{abcde} \left( A^{e}_{c} B^{ab}_{de} - A^{a}_{e} B^{be}_{cd} \right) \lambda^{ab}_{cd}
\end{align}

\begin{align}
    C^a_b &= \sum_{cd} (n_c - n_d)\, A^{c}_{d} B^{ad}_{bc}
\end{align}

\begin{align}
    C^{ab}_{cd} &= 2 \sum_{e} \left( A^{e}_{c} B^{ab}_{de} - A^{a}_{e} B^{be}_{cd} \right)
\end{align}

\item \textbf{The 13X commutator}

\begin{align}
    C_0 = \frac{1}{12} \sum_{abcdefg}(A^{a}_ {g}B^{b c g}_{d e f} - A^{g}_ {d}B^{a b c}_{e f g})\lambda^{a b c}_ {d e f}
\end{align}

\begin{align}
    C^a_b = \frac{1}{2}\sum_{cdefg} ( A^{g}_ {e}B^{a c d}_ {b f g} -A^{c}_ {g}B^{a d g}_ {b e f} )\lambda^{c d}_ {e f}
\end{align}

\begin{align}
    C^{ab}_{cd} = \sum_{ef}(n^{}_ {e} - n^{}_ {f})A^{e}_ {f}B^{a b f}_ {c d e} 
\end{align}

\begin{align}
    C^{abc}_{def} = 3\sum_{g } ( A^{a}_ {g}B^{b c g}_ {d e f} - A^{g}_ {d}B^{a b c}_ {e f g} ) 
\end{align}

\item \textbf{The 22X commutator}

\begin{equation}
\begin{aligned}
     C_0 &= \frac{1}{4}\sum_{abcd}(n_a n_b \bar{n}_c \bar{n}_d - \bar{n}_a \bar{n}_b n_c n_d)A^{a b}_{c d}B^{c d}_{a b} \\
    &+ \sum_{abcdef} \Big[(n_{e} - n_{f})A^{a e}_{c f}B^{b f}_{d e} \\
    &+\frac{1}{8} (n_e n_f - \bar{n}_e \bar{n}_f)(A^{e f}_{c d}B^{a b}_{e f} - A^{a b}_{e f}B^{e f}_{c d}) \Big]\lambda^{a b}_ {c d} \\
    &+ \frac{1}{4}\sum_{abcdefg} ( A^{a g}_{d e}B^{b c}_{f g} - A^{a b}_{d g}B^{c g}_{e f} )\lambda^{a b c}_{d e f} \\
\end{aligned}
\end{equation}

\begin{equation}
\begin{aligned}
    C^a_b &= \frac{1}{2} \sum_{cde}(n_c \bar{n}_d \bar{n}_e + \bar{n}_c n_d n_e )(A^{ac}_{de}B^{de}_{bc} - A^{de}_{bc}B^{ac}_{de})  \\
    &+  \sum_{cdefg} \Big[ \frac{1}{4}(A^{ag}_{ef}B^{cd}_{bg} - A^{cd}_{bg}B^{ag}_{ef}) + (A^{ac}_{eg}B^{dg}_{bf} - A^{cg}_{be}B^{ad}_{fg}) \\
    &+ \frac{1}{2}(A^{ag}_{be}B^{cd}_{fg} - A^{ac}_{bg}B^{dg}_{ef}) + \frac{1}{2}(A^{cd}_{eg}B^{ag}_{bf} - A^{cg}_{ef}B^{ad}_{bg}) \Big]\lambda^{c d}_{ef}
\end{aligned}
\end{equation}

\begin{equation}
\begin{aligned}
    C^{ab}_{cd} &= \sum_{ef}  4(n_{e} - n_{f} )A^{a e}_ {c f}B^{b f}_ {d e} \\
    & + \frac{1}{2} \sum_{ef} (n_e n_f - \bar{n}_e \bar{n}_f )(A^{e f}_ {c d}B^{a b}_ {e f} - A^{a b}_ {e f}B^{e f}_ {c d}) 
\end{aligned}
\end{equation}

\begin{equation}
\begin{aligned}
    C^{abc}_{def} = \sum_{g} 9( A^{a g}_ {d e}B^{b c}_ {f g} -A^{a b}_ {d g}B^{c g}_ {e f} ) 
\end{aligned}
\end{equation}

\item \textbf{The 23X commutator}

\begin{equation}
\begin{aligned}
    C_0 &= \frac{1}{4}\sum_{abcd}\sum_{efg}(n_e \bar{n}_f \bar{n}_g + \bar{n}_e n_f n_g )(A^{fg}_{ce} B^{abe}_{dfg} - A^{ae}_{fg} B^{bfg}_{cde}) \lambda^{ab}_{cd} \\
    &+ \frac{1}{8} \sum_{abcd} \sum_{efgh} \sum_{i}(A^{a b}_ {g i}B^{e f i}_ {c d h} -A^{ei}_{cd}B^{abf}_{ghi} \\
    &+ 4A^{a i}_ {c g}B^{b e f}_ {d h i}  -4A^{ae}_{ci}B^{bfi}_{dgh} )\lambda^{a b}_ {c d}\lambda^{e f}_ {g h} \\
    &+ \sum_{abcdef} \sum_{gh} \Big[\frac{1}{4}(n^{}_ {g} - n^{}_ {h})A^{a g}_ {d h}B^{b c h}_ {e f g}  \\
    &+ \frac{1}{24}(n_g n_h -\bar{n}_g \bar{n}_h)(A^{g h}_ {d e}B^{a b c}_ {f g h} - A^{a b}_ {g h}B^{c g h}_ {d e f})\Big]\lambda^{a b c}_ {d e f} \\
    &+ \sum_{abcdefghi} \frac{1}{24}(A^{a b}_ {e i}B^{c d i}_ {f g h} - A^{a i}_ {e f}B^{b c d}_ {g h i})\lambda^{a b c d}_ {e f g h} 
\end{aligned}
\end{equation}

\begin{equation}
\begin{aligned}
    C^a_b &=  \frac{1}{4}\sum_{cdef}(n_c n_d \bar{n}_e \bar{n}_f - \bar{n}_c \bar{n}_d n_e n_f)A^{c d}_ {e f}B^{a e f}_ {b c d} \\
    &+ \sum_{cdefgh}\Big[ (n^{}_ {g} - n^{}_ {h})A^{c g}_ {e h}B^{a d h}_ {b f g} \\
    &-\frac{1}{2}(n^{}_ {g} - n^{}_ {h})(A^{a g}_ {e h}B^{c d h}_ {b f g} + A^{c g}_ {b h}B^{a d h}_ {e f g})  \\
    &+ \frac{1}{4} (n_g n_h - \bar{n}_g \bar{n}_h)(A^{g h}_ {b e}B^{a c d}_ {f g h} - A^{a c}_ {g h}B^{d g h}_ {b e f}) \\
    &+ \frac{1}{8}(n_g n_h - \bar{n}_g \bar{n}_h)(A^{g h}_ {e f}B^{a c d}_ {b g h} - A^{c d}_ {g h}B^{a g h}_ {b e f}) \Big] \lambda^{c d}_ {e f}  \\
    &+ \sum_{cdefghi} \Big[ \frac{1}{4}( A^{c i}_ {b f}B^{a d e}_ {g h i}-A^{a c}_ {f i}B^{d e i}_ {b g h})  + \frac{1}{4}(A^{c i}_ {f g}B^{a d e}_ {b h i}- A^{c d}_ {f i}B^{a e i}_ {b g h}) \\
    &+ \frac{1}{12}(A^{a c}_ {b i}B^{d e i}_ {f g h}- A^{a i}_ {b f}B^{c d e}_ {g h i}) +\frac{1}{12}( A^{c d}_ {b i}B^{a e i}_ {f g h} - A^{a i}_ {f g}B^{c d e}_ {b h i} )  \Big] \lambda^{c d e}_ {f g h} 
\end{aligned}
\end{equation}

\begin{equation}
\begin{aligned}
    C^{ab}_{cd} &=  \sum_{efg}(n_e \bar{n}_f \bar{n}_g + \bar{n}_e n_f n_g)(A^{fg}_{ce}B^{abe}_{dfg} - A^{ae}_{fg}B^{bfg}_{cde})\\
    &+  \sum_{efghi} \Big[ \frac{1}{2}( A^{ab}_{gi} B^{efi}_{cdh} - A^{ei}_{cd} B^{abf}_{ghi}) \\
    &+ \frac{1}{2}( A^{ef}_{ci} B^{abi}_{dgh} - A^{ai}_{gh} B^{bef}_{cdi}) + \frac{1}{2}( A^{ef}_{gi} B^{abi}_{cdh} - A^{ei}_{gh} B^{abf}_{cdi}) \\
    &+ 2( A^{ei}_{cg} B^{abf}_{dhi} - A^{ae}_{gi} B^{bfi}_{cdh})+ 2( A^{ai}_{cg} B^{bef}_{dhi} - A^{ae}_{ci} B^{bfi}_{dgh})\Big] \lambda^{ef}_{gh}
\end{aligned}
\end{equation}

\begin{equation}
\begin{aligned}
    C^{abc}_{def} &= 9\sum_{gh} (n_g - n_h)A^{ag}_{dh} B^{bch}_{efg} \\
    &+ \frac{3}{2}\sum_{gh} (n_g n_h - \bar{n}_g \bar{n}_h)(A^{gh}_{de} B^{abc}_{fgh} - A^{ab}_{gh} B^{cgh}_{def}) 
\end{aligned}
\end{equation}

\item \textbf{The 33X commutator:}
Its contribution to the zero-body piece reads,
\begin{equation}
\begin{aligned} 
    C_0 &= \frac{1}{36} \sum_{abcdef}(n_a n_b n_c \bar{n}_d \bar{n}_e \bar{n}_f - \bar{n}_a \bar{n}_b \bar{n}_c n_d n_e n_f  )A^{abc}_{def}B^{def}_{abc} \\
    &+  \sum_{abcd} \sum_{efgh}\Big[ \frac{1}{4}(n_e n_f \bar{n}_g \bar{n}_h - \bar{n}_e \bar{n}_f n_g n_h)A^{aef}_{cgh} B^{bgh}_{def} \\
    &+ \frac{1}{24}(n_e \bar{n}_f \bar{n}_g \bar{n}_h - \bar{n}_e n_f n_g n_h )(A^{abe}_{fgh} B^{fgh}_{cde} - A^{fgh}_{cde} B^{abe}_{fgh}) \Big] \lambda^{ab}_{cd} \\
    &+ \sum_{abcdef}\sum_{ghi} \Big[ \frac{1}{8}(n_g \bar{n}_h \bar{n}_i + \bar{n}_g n_h n_i)(A^{ahi}_{deg}B^{bcg}_{fhi} - A^{abg}_{dhi}B^{chi}_{efg}) \\
    &+ \frac{1}{216} (n_g n_h n_i + \bar{n}_g \bar{n}_h \bar{n}_i)(A^{abc}_{ghi}B^{ghi}_{def} - A^{ghi}_{def}B^{abc}_{ghi}) \Big]\lambda^{abc}_{def} \\
    &+ \sum_{abcdefgh}\sum_{ij} \Big[\frac{1}{16} (n_i - n_j)A^{abi}_{efj}B^{cdj}_{ghi} \\
    &+ \frac{1}{72} (n_i n_j - \bar{n}_i \bar{n}_j)(A^{aij}_{efg}B^{bcd}_{hij} - A^{abc}_{eij}B^{dij}_{fgh}) \Big] \lambda^{abcd}_{efgh} \\
      \nonumber
    \end{aligned}
\end{equation}
\begin{equation}
\begin{aligned}
    &+ \sum_{abcd} \sum_{efgh}\sum_{ij} \Big[ \frac{1}{16} (n_i - n_j)A^{abi}_{ghj} B^{efj}_{cdi}  \\ 
    &- \frac{1}{4} (n_i - n_j)(A^{abi}_{cgj} B^{efj}_{dhi}+A^{aei}_{cdj} B^{bfj}_{ghi}) \\
    &+ \frac{1}{2} (n_i - n_j)A^{aei}_{cgj} B^{bfj}_{dhi} \Big] \lambda^{ab}_{cd}\lambda^{ef}_{gh}\\
    &+\sum_{abcdefgh}\sum_{ij} \Big[ \frac{1}{8}(n_i n_j - \bar{n}_i \bar{n}_j)(A^{aij}_{cdg}B^{bef}_{hij}  +A^{aij}_{cgh}B^{bef}_{dij} \\
    &-A^{abe}_{cij}B^{fij}_{dgh} - A^{abe}_{gij}B^{fij}_{cdh}) \Big] \lambda^{ab}_{cd}\lambda^{ef}_{gh} \\
    &+ \sum_{abcdefghij}\sum_k\frac{1}{144}(A^{abc}_{fgk}B^{dek}_{hij} - A^{abk}_{fgh}B^{cde}_{ijk})\lambda^{abcde}_{fghij}  \nonumber
    \end{aligned}
\end{equation}
\begin{equation}
\begin{aligned}
    &+ \sum_{abcd}\sum_{efghij}\sum_{k} \Big[ \frac{1}{4}(A^{a e k}_ {c h i}B^{b f g}_ {d j k} - A^{a e f}_ {c h k}B^{b g k}_ {d i j}) \\
    &+ \frac{1}{8}( A^{a e f}_ {h i k}B^{b g k}_ {c d j} + A^{a e k}_ {c d h}B^{b f g}_ {i j k} -A^{a b e}_ {c h k}B^{f g k}_ {d i j} - A^{e f k}_ {c h i}B^{a b g}_ {d j k}) \\
    &+ \frac{1}{16}(A^{a b e}_ {h i k}B^{f g k}_ {c d j} - A^{e f k}_ {c d h}B^{a b g}_ {i j k})\\
    &+\frac{1}{24}( A^{a e f}_ {c d k}B^{b g k}_ {h i j}   + A^{a e k}_ {h i j}B^{b f g}_ {c d k} - A^{a b k}_ {c h i}B^{e f g}_ {d j k} - A^{e f g}_ {c h k}B^{a b k}_ {d i j}  ) \\
    &+\frac{1}{144}( A^{e f g}_ {c d k}B^{a b k}_ {h i j} - A^{a b k}_ {h i j}B^{e f g}_ {c d k} ) \Big]
    \lambda^{a b}_ {c d}
    \lambda^{e f g}_ {h i j}
\end{aligned}
\end{equation}
Its contribution to the one-body piece reads,
\begin{equation}
\begin{aligned}
    C^{a}_{b} &=  \frac{1}{12}\sum_{cdefg} (n_c n_d n_e \bar{n}_f \bar{n}_g + \bar{n}_c \bar{n}_d \bar{n}_e n_f n_g)(A^{afg}_{cde}B^{cde}_{bfg} - A^{cde}_{bfg}B^{afg}_{cde})  \\
    &+ \sum_{cdef} \sum_{ghi} \Big[ \frac{1}{2}(n_g \bar{n}_h \bar{n}_i + \bar{n}_g n_h n_i)(A^{acg}_{ehi}B^{dhi}_{bfg} - A^{chi}_{beg}B^{adg}_{fhi}) \\
    &+ \frac{1}{4}(n_g \bar{n}_h \bar{n}_i + \bar{n}_g n_h n_i)((A^{ahi}_{beg}B^{cdg}_{fhi} - A^{acg}_{bhi}B^{dhi}_{efg}) \\
    &+ (A^{cdg}_{ehi}B^{ahi}_{bfg} - A^{chi}_{efg}B^{adg}_{bhi})) \\
    &+ \frac{1}{8} (n_g \bar{n}_h \bar{n}_i + \bar{n}_g n_h n_i)(A^{ahi}_{efg}B^{cdg}_{bhi} - A^{cdg}_{bhi}B^{ahi}_{efg}) \\
    &+ \frac{1}{24}(n_g n_h n_i + \bar{n}_g \bar{n}_h \bar{n}_i)(A^{acd}_{ghi}B^{ghi}_{bef} - A^{ghi}_{bef}B^{acd}_{ghi})\Big] \lambda^{cd}_{ef} 
    \nonumber
    \end{aligned}
\end{equation}
\begin{equation}
    \begin{aligned}
    &+  \sum_{cdefgh} \sum_{ij} \Big[ \frac{1}{4} (n_i - n_j)((A^{aci}_{bfj}B^{dej}_{ghi}+A^{cdi}_{fgj}B^{aej}_{bhi}) \\
    &+(A^{aci}_{fgj}B^{dej}_{bhi}+A^{cdi}_{bfj}B^{aej}_{ghi})) \\
    &+ \frac{1}{8} (n_i n_j - \bar{n}_i \bar{n}_j)(A^{acd}_{fij}B^{eij}_{bgh} - A^{cij}_{bfg}B^{ade}_{hij}) \\
    &+ \frac{1}{24} (n_i n_j - \bar{n}_i \bar{n}_j)\Big((A^{aij}_{bfg}B^{cde}_{hij} - A^{acd}_{bij}B^{eij}_{fgh}) \\
    &+ (A^{cij}_{fgh}B^{ade}_{bij} - A^{cde}_{fij}B^{aij}_{bgh})\Big) \\
    &+ \frac{1}{72} (n_i n_j - \bar{n}_i \bar{n}_j)(A^{cde}_{bij}B^{aij}_{fgh} - A^{aij}_{fgh}B^{cde}_{bij}) \Big] \lambda^{cde}_{fgh} 
    \nonumber
    \end{aligned}
\end{equation}
\begin{equation}
    \begin{aligned}
    &+  \sum_{cdefghij} \sum_{k} \Big[ \frac{1}{16}(A^{a c d}_ {g h k}B^{e f k}_ {b i j}- A^{c d k}_ {b g h}B^{a e f}_ {i j k})\\
    &+\frac{1}{24}(A^{a c d}_ {b g k}B^{e f k}_ {h i j} - A^{a c k}_ {b g h}B^{d e f}_ {i j k} + A^{c d k}_ {g h i}B^{a e f}_ {b j k}- A^{c d e}_ {g h k}B^{a f k}_ {b i j}) \\
    &+ \frac{1}{36}(  A^{a c k}_ {g h i}B^{d e f}_ {b j k} - A^{c d e}_ {b g k}B^{a f k}_ {h i j} )  \Big] \lambda^{cdef}_{ghij} \\
 &+\sum_{cdef}\sum_{ghij}\sum_{k} \Big[ \frac{1}{2}\Big( (A^{a c g}_ {e i k}B^{d h k}_ {b f j} - A^{c g k}_ {b e i}B^{a d h}_ {f j k}) \\
    &+ (A^{a c k}_ {b e i}B^{d g h}_ {f j k} - A^{a c g}_ {b e k}B^{d h k}_ {f i j}) + (A^{c d g}_ {e i k}B^{a h k}_ {b f j}-A^{c g k}_ {e i j}B^{a d h}_ {b f k}) \Big) \\ 
    &+ \frac{1}{4}\Big( (A^{a c k}_ {e f i}B^{d g h}_ {b j k}- A^{c g h}_ {b i k}B^{a d k}_ {e f j} ) + (A^{a c k}_ {e i j}B^{d g h}_ {b f k}- A^{c d g}_ {b i k}B^{a h k}_ {e f j}) \\
    &+ (A^{c d k}_ {b e i}B^{a g h}_ {f j k} -A^{a c g}_ {i j k}B^{d h k}_ {b e f}) + (A^{c g k}_ {b e f}B^{a d h}_ {i j k}  - A^{a c d}_ {e i k}B^{g h k}_ {b f j}) 
    \Big) \\
    &+ \frac{1}{8}\Big((A^{a c d}_ {b i k}B^{g h k}_ {e f j} - A^{a c k}_ {b i j}B^{d g h}_ {e f k}) + (A^{c d k}_ {e i j}B^{a g h}_ {b f k}   - A^{c d g}_ {i j k}B^{a h k}_ {b e f})\Big) \\
    &+ \frac{1}{16}(A^{a c d}_ {i j k}B^{g h k}_ {b e f} - A^{c d k}_ {b i j}B^{a g h}_ {e f k})   
     \Big] \lambda^{cd}_{ef}\lambda^{gh}_{ij} 
\end{aligned}
\end{equation}
Its contribution to the two-body piece reads,
\begin{equation}
\begin{aligned}
    C^{ab}_{cd} &= \sum_{efgh} \Big[ (n_e n_f \bar{n}_g \bar{n}_h - \bar{n}_e \bar{n}_f n_g n_h)A^{aef}_{cgh}B^{bgh}_{def} \\
    &+ \frac{1}{6}(n_e \bar{n}_f \bar{n}_g \bar{n}_h - \bar{n}_e n_f n_g n_h)(A^{abe}_{fgh}B^{fgh}_{cde} - A^{fgh}_{cde}B^{abe}_{fgh}) \Big]  \\
    &+\sum_{efgh} \sum_{ij} \Big[ 4(n_i - n_j)A^{aei}_{cgj}B^{bfj}_{dhi} \\
    &- (n_i - n_j)\Big((A^{abi}_{cgj}B^{efj}_{dhi}+A^{aei}_{cdj}B^{bfj}_{ghi})+(A^{efi}_{cgj}B^{abj}_{dhi}+A^{aei}_{ghj}B^{bfj}_{cdi})\Big) \\
    &+ \frac{1}{4}(n_i - n_j)(A^{abi}_{ghj}B^{efj}_{cdi} + A^{efi}_{cdj}B^{abj}_{ghi} )+ \frac{1}{2}(n_i n_j - \bar{n}_i \bar{n}_j)\\
    &\Big((A^{eij}_{cdg}B^{abf}_{hij} - A^{abe}_{gij}B^{fij}_{cdh})  + (A^{eij}_{cgh}B^{abf}_{dij} - A^{aef}_{gij}B^{bij}_{cdh} ) \\
    &+ (A^{aij}_{cdg}B^{bef}_{hij} - A^{abe}_{cij}B^{fij}_{dgh} ) + (A^{aij}_{cgh}B^{bef}_{dij} - A^{aef}_{cij}B^{bij}_{dgh}) \Big) \Big] \lambda^{ef}_{gh} \\
    &+ \sum_{efghij} \sum_{k} \Big[ (A^{a e k}_ {c h i}B^{b f g}_ {d j k}- A^{a e f}_ {c h k}B^{b g k}_ {d i j})\\ 
    &+ \frac{1}{2}\Big((A^{a e f}_ {h i k}B^{b g k}_ {c d j} - A^{e f k}_ {c h i}B^{a b g}_ {d j k})+( A^{a e k}_ {c d h}B^{b f g}_ {i j k} - A^{a b e}_ {c h k}B^{f g k}_ {d i j} )\Big) \\
    &+ \frac{1}{4}(A^{a b e}_ {h i k}B^{f g k}_ {c d j} - A^{e f k}_ {c d h}B^{a b g}_ {i j k} ) \\
    &+ \frac{1}{6}\Big((A^{a e f}_ {c d k}B^{b g k}_ {h i j} - A^{a b k}_ {c h i}B^{e f g}_ {d j k}) + (A^{a e k}_ {h i j}B^{b f g}_ {c d k} - A^{e f g}_ {c h k}B^{a b k}_ {d i j} )\Big)\\ 
    &+ \frac{1}{12}\Big((A^{a b e}_ {c d k}B^{f g k}_ {h i j} - A^{a b k}_ {c d h}B^{e f g}_ {i j k}) + (A^{e f g}_ {h i k}B^{a b k}_ {c d j}  - A^{e f k}_ {h i j}B^{a b g}_ {c d k} )\Big) \\
    &+ \frac{1}{36}(A^{e f g}_ {c d k}B^{a b k}_ {h i j} - A^{a b k}_ {h i j}B^{e f g}_ {c d k}  ) \Big] \lambda^{efg}_{hij} 
\end{aligned}
\end{equation}
Its contribution to the three-body piece reads,
\begin{equation}
\begin{aligned}
    C^{abc}_{def} &= \sum_{ghi} \Big[ \frac{9}{2}(n_g \bar{n}_h \bar{n}_i + \bar{n}_g n_h n_i)(A^{ahi}_{deg}B^{bcg}_{fhi} - A^{abg}_{dhi}B^{chi}_{efg} ) \\
    &+ \frac{1}{6} (n_g n_h n_i + \bar{n}_g \bar{n}_h \bar{n}_i)(A^{abc}_{ghi}B^{ghi}_{def} - A^{ghi}_{def}B^{abc}_{ghi}) \Big] \\
    &+ \sum_{ghij} \sum_{k} \Big[ 9(A^{a g k}_ {d e i}B^{b c h}_ {f j k} - A^{a b g}_ {d i k}B^{c h k}_ {e f j} ) \\
    &+ \frac{9}{2}\Big((A^{a b k}_ {d e i}B^{c g h}_ {f j k} - A^{a b g}_ {d e k}B^{c h k}_ {f i j}) + (A^{a g h}_ {d i k}B^{b c k}_ {e f j}  - A^{a g k}_ {d i j}B^{b c h}_ {e f k} )\Big) \\
    &+ \frac{9}{4}(A^{a g h}_ {d e k}B^{b c k}_ {f i j}- A^{a b k}_ {d i j}B^{c g h}_ {e f k})\\
    &+ \frac{3}{2}\Big((A^{a b c}_ {d i k}B^{g h k}_ {e f j} - A^{a g k}_ {d e f}B^{b c h}_ {i j k})+ (A^{g h k}_ {d e i}B^{a b c}_ {f j k} - A^{a b g}_ {i j k}B^{c h k}_ {d e f} )\Big) \\
    &+ \frac{1}{4}(A^{a b c}_ {i j k}B^{g h k}_ {d e f}    - A^{g h k}_ {d e f}B^{a b c}_ {i j k}) \Big] \lambda^{gh}_{ij}
\end{aligned}
\end{equation}

\end{itemize} 

\section{The MR-IMSRG(3) flow}
\label{app:IMSRG3}

In this appendix, we present the explicit expressions for all terms entering the MR-IMSRG(3) flow equations.

\begin{itemize}
\item  \textbf{Zero-body flow equation:} The zero-body flow is given by Eq.~(\ref{eq:IMSRG3-0b}). When the
$\lambda^{(2)}\lambda^{(2)}$ contribution is reconstructed from the two-body flow, the following part of
$d\Gamma^{ab}_{cd}(23,\lambda^{(2)})$ must be subtracted:
\begin{equation}
\label{eq:dGamma23-non}
\begin{aligned}
&d\Gamma^{ab}_{cd}
 (23,\lambda^{(2)},{\rm non})
\\
&=
\frac{1}{2}\sum_{efghi}
\left(
\eta^{ef}_{gi}W^{abi}_{cdh}
-\eta^{ei}_{gh}W^{abf}_{cdi}
\right)\lambda^{ef}_{gh}
\\
&-\frac{1}{2}\sum_{efghi}
\left(
\Gamma^{ef}_{gi}\eta^{abi}_{cdh}
-\Gamma^{ei}_{gh}\eta^{abf}_{cdi}
\right)\lambda^{ef}_{gh}.
\end{aligned}
\end{equation}

The remaining zero-body terms involving four- and five-body
irreducible density matrices are
\begin{equation}
\label{eq:dE23-lambda4}
\begin{aligned}
dE(23,\lambda^{(4)})
&=
\frac{1}{24}\sum_{abcdefghi}
\left(
\eta^{ab}_{ei}W^{cdi}_{fgh}
-\eta^{ai}_{ef}W^{bcd}_{ghi}
\right)
\lambda^{abcd}_{efgh}
\\
&-\frac{1}{24}\sum_{abcdefghi}
\left(
\Gamma^{ab}_{ei}\eta^{cdi}_{fgh}
-\Gamma^{ai}_{ef}\eta^{bcd}_{ghi}
\right)
\lambda^{abcd}_{efgh},
\end{aligned}
\end{equation}

\begin{equation}
\label{eq:dE33-lambda4}
\begin{aligned}
dE(33,\lambda^{(4)})
&=
\sum_{abcdefgh}\sum_{ij}
\Bigg[
\frac{1}{16}(n_i-n_j)
\eta^{abi}_{efj}W^{cdj}_{ghi}
\\
&+ \frac{1}{72}
(n_i n_j-\bar n_i\bar n_j)
\left(
\eta^{aij}_{efg}W^{bcd}_{hij}
-\eta^{abc}_{eij}W^{dij}_{fgh}
\right)
\Bigg]\lambda^{abcd}_{efgh},
\end{aligned}
\end{equation}

and
\begin{equation}
\label{eq:dE33-lambda5}
\begin{aligned}
dE(33,\lambda^{(5)})
=&
\frac{1}{144}
\sum_{abcdefghijk}
\left(
\eta^{abc}_{fgk}W^{dek}_{hij}
-\eta^{abk}_{fgh}W^{cde}_{ijk}
\right)
\lambda^{abcde}_{fghij}.
\end{aligned}
\end{equation}

\item \textbf{One-body flow equation:}
The explicit expressions for all terms appearing on the right-hand side of Eq.~(\ref{eq:IMSRG3-1b}) are given below,

\begin{align}
    df^{a}_{b}(11) &= \sum_{c} (\eta^{a}_ {c}f^{c}_ {b} - \eta^{c}_ {b}f^{a}_ {c}), 
\end{align}

\begin{align}
    df^{a}_{b}(12) &= \sum_{cd} (n^{}_ {c} - n^{}_ {d})\eta^{c}_{d}\Gamma^{a d}_ {b c}  - \sum_{cd} (n^{}_ {c} - n^{}_ {d})f^{c}_{d}\eta^{a d}_ {b c},
\end{align}

\begin{align}
    df^{a}_{b}(13) &= \sum_{cdefg} \frac{1}{2}( \eta^{g}_ {e}W^{a c d}_ {b f g} -\eta^{c}_ {g}W^{a d g}_ {b e f} )\lambda^{c d}_ {e f} \nonumber\\
    &- \sum_{cdefg} \frac{1}{2}( f^{g}_ {e}\eta^{a c d}_ {b f g} -f^{c}_ {g}\eta^{a d g}_ {b e f} )\lambda^{c d}_ {e f}, 
\end{align}

\begin{align}
    df^{a}_{b}(22) 
     &= \frac{1}{2} \sum_{cde}(n_c \bar{n}_d \bar{n}_e + \bar{n}_c n_d n_e )(\eta^{ac}_{de}\Gamma^{de}_{bc} - \eta^{de}_{bc}\Gamma^{ac}_{de}) \nonumber \\
    &+  \sum_{cdefg} \Big[ \frac{1}{4}(\eta^{ag}_{ef}\Gamma^{cd}_{bg} - \eta^{cd}_{bg}\Gamma^{ag}_{ef}) + (\eta^{ac}_{eg}\Gamma^{dg}_{bf} - \eta^{cg}_{be}\Gamma^{ad}_{fg}) \nonumber\\
    &+ \frac{1}{2}(\eta^{ag}_{be}\Gamma^{cd}_{fg} - \eta^{ac}_{bg}\Gamma^{dg}_{ef}) + \frac{1}{2}(\eta^{cd}_{eg}\Gamma^{ag}_{bf} - \eta^{cg}_{ef}\Gamma^{ad}_{bg}) \Big]\lambda^{c d}_{ef}.
\end{align}

The $df^{a}_{b}(23)$ term is composed of three terms,
\begin{equation}
    df^{a}_{b}(23) = df^{a}_{b}(23,\lambda^{(1)}) + df^{a}_{b}(23,\lambda^{(2)}) + df^{a}_{b}(23,\lambda^{(3)}),
\end{equation}
which are obtained as
\begin{equation}
\begin{aligned}
    &df^{a}_{b} (23,\lambda^{(1)}) \\
    &= \frac{1}{4}\sum_{cdef}(n_c n_d \bar{n}_e \bar{n}_f - \bar{n}_c \bar{n}_d n_e n_f)\eta^{c d}_ {e f}W^{a e f}_ {b c d} ,\\
    &-  \sum_{cdef}\frac{1}{4}(n_c n_d \bar{n}_e \bar{n}_f - \bar{n}_c \bar{n}_d n_e n_f)\Gamma^{c d}_ {e f}\eta^{a e f}_ {b c d}  \nonumber\\
\end{aligned}
\end{equation}

\begin{align}
    df^{a}_{b}(23,\lambda^{{(2)}}) 
     =&   \sum_{cdefgh}\Big[ (n^{}_ {g} - n^{}_ {h})\eta^{c g}_ {e h}W^{a d h}_ {b f g} \nonumber\\ 
    &-\frac{1}{2}(n^{}_ {g} - n^{}_ {h})(\eta^{a g}_ {e h}W^{c d h}_ {b f g} + \eta^{c g}_ {b h}W^{a d h}_ {e f g})  \nonumber\\
    &+ \frac{1}{4} (n_g n_h - \bar{n}_g \bar{n}_h)(\eta^{g h}_ {b e}W^{a c d}_ {f g h} - \eta^{a c}_ {g h}W^{d g h}_ {b e f})  \nonumber\\
    &+ \frac{1}{8}(n_g n_h - \bar{n}_g \bar{n}_h)(\eta^{g h}_ {e f}W^{a c d}_ {b g h} - \eta^{c d}_ {g h}W^{a g h}_ {b e f}) \Big] \lambda^{c d}_ {e f}  \nonumber\\
    &- \sum_{cdefgh}\Big[ (n^{}_ {g} - n^{}_ {h})\Gamma^{c g}_ {e h}\eta^{a d h}_ {b f g} \nonumber\\
    &-\frac{1}{2}(n^{}_ {g} - n^{}_ {h})(\Gamma^{a g}_ {e h}\eta^{c d h}_ {b f g} + \Gamma^{c g}_ {b h}\eta^{a d h}_ {e f g}) \nonumber \\
    &+ \frac{1}{4} (n_g n_h - \bar{n}_g \bar{n}_h)(\Gamma^{g h}_ {b e}\eta^{a c d}_ {f g h} - \Gamma^{a c}_ {g h}\eta^{d g h}_ {b e f}) \nonumber\\ 
    &+ \frac{1}{8}(n_g n_h - \bar{n}_g \bar{n}_h)(\Gamma^{g h}_ {e f}\eta^{a c d}_ {b g h} - \Gamma^{c d}_ {g h}\eta^{a g h}_ {b e f}) \Big] \lambda^{c d}_ {e f},
\end{align}

\begin{align}
&df^a_b(23,\lambda^{(3)}) \nonumber  \\
&=
\sum_{cdefghi}
\Bigg[
\frac{1}{4}
\left(
\eta^{ci}_{bf}W^{ade}_{ghi}
-\eta^{ac}_{fi}W^{dei}_{bgh}
\right)
+\frac{1}{4}
\left(
\eta^{ci}_{fg}W^{ade}_{bhi}
-\eta^{cd}_{fi}W^{aei}_{bgh}
\right)
\nonumber\\
&\qquad
+\frac{1}{12}
\left(
\eta^{ac}_{bi}W^{dei}_{fgh}
-\eta^{ai}_{bf}W^{cde}_{ghi}
\right)
+\frac{1}{12}
\left(
\eta^{cd}_{bi}W^{aei}_{fgh}
-\eta^{ai}_{fg}W^{cde}_{bhi}
\right)
\Bigg]\lambda^{cde}_{fgh}
\nonumber\\
&-
\sum_{cdefghi}
\Bigg[
\frac{1}{4}
\left(
\Gamma^{ci}_{bf}\eta^{ade}_{ghi}
-\Gamma^{ac}_{fi}\eta^{dei}_{bgh}
\right)
+\frac{1}{4}
\left(
\Gamma^{ci}_{fg}\eta^{ade}_{bhi}
-\Gamma^{cd}_{fi}\eta^{aei}_{bgh}
\right)
\nonumber\\
&\qquad
+\frac{1}{12}
\left(
\Gamma^{ac}_{bi}\eta^{dei}_{fgh}
-\Gamma^{ai}_{bf}\eta^{cde}_{ghi}
\right)
+\frac{1}{12}
\left(
\Gamma^{cd}_{bi}\eta^{aei}_{fgh}
-\Gamma^{ai}_{fg}\eta^{cde}_{bhi}
\right)
\Bigg]\lambda^{cde}_{fgh}.
\label{eq:df23-lambda3}
\end{align}

The $df^{a}_{b}(33)$ term is composed of five terms, 
\begin{align}
    df^a_b(33)
=&df^a_b(33,\lambda^{(1)}) + df^a_b(33,\lambda^{(2)}) +df^a_b(33,\lambda^{(3)}) \nonumber \\
&+df^a_b(33,\lambda^{(4)})
+df^a_b(33,\lambda^{(2)}\lambda^{(2)}),
\end{align} 

which are obtained as 
\begin{align}
    df^a_b(33,\lambda^{(1)}) &= \frac{1}{12}\sum_{cdefg} (n_c n_d n_e \bar{n}_f \bar{n}_g + \bar{n}_c \bar{n}_d \bar{n}_e n_f n_g)\nonumber\\
    &\times(\eta^{afg}_{cde}W^{cde}_{bfg} - \eta^{cde}_{bfg}W^{afg}_{cde}),
\end{align} 

\begin{equation}
\begin{aligned}
    &df^a_b(33,\lambda^{(2)}) \\ 
     &= \sum_{cdef} \sum_{ghi} \Big[ \frac{1}{2}(n_g \bar{n}_h \bar{n}_i + \bar{n}_g n_h n_i)(\eta^{acg}_{ehi}W^{dhi}_{bfg} - \eta^{chi}_{beg}W^{adg}_{fhi}) \\
    &+ \frac{1}{4}(n_g \bar{n}_h \bar{n}_i + \bar{n}_g n_h n_i)((\eta^{ahi}_{beg}W^{cdg}_{fhi} - \eta^{acg}_{bhi}W^{dhi}_{efg}) \\
    &+ (\eta^{cdg}_{ehi}W^{ahi}_{bfg} - \eta^{chi}_{efg}W^{adg}_{bhi})) \\
    &+ \frac{1}{8} (n_g \bar{n}_h \bar{n}_i + \bar{n}_g n_h n_i)(\eta^{ahi}_{efg}W^{cdg}_{bhi} - \eta^{cdg}_{bhi}W^{ahi}_{efg}) \\
    &+ \frac{1}{24}(n_g n_h n_i + \bar{n}_g \bar{n}_h \bar{n}_i)(\eta^{acd}_{ghi}W^{ghi}_{bef} - \eta^{ghi}_{bef}W^{acd}_{ghi})\Big] \lambda^{cd}_{ef} ,
\end{aligned}
\end{equation}

\begin{align}
&df^a_b(33,\lambda^{(3)}) \nonumber \\
=&
\sum_{cdefgh}\sum_{ij}
\Bigg[
\frac{1}{4}(n_i-n_j)
\Big(
\eta^{aci}_{bfj}W^{dej}_{ghi}
+\eta^{cdi}_{fgj}W^{aej}_{bhi}
\nonumber\\
&
+\eta^{aci}_{fgj}W^{dej}_{bhi}
+\eta^{cdi}_{bfj}W^{aej}_{ghi}
\Big)
\nonumber\\
&
+\frac{1}{8}
(n_i n_j-\bar n_i\bar n_j)
\left(
\eta^{acd}_{fij}W^{eij}_{bgh}
-\eta^{cij}_{bfg}W^{ade}_{hij}
\right)
\nonumber\\
&
+\frac{1}{24}
(n_i n_j-\bar n_i\bar n_j)
\Big[
\eta^{aij}_{bfg}W^{cde}_{hij}
-\eta^{acd}_{bij}W^{eij}_{fgh}
\nonumber\\
&
+\eta^{cij}_{fgh}W^{ade}_{bij}
-\eta^{cde}_{fij}W^{aij}_{bgh}
\Big]
\nonumber\\
&
+\frac{1}{72}
(n_i n_j-\bar n_i\bar n_j)
\left(
\eta^{cde}_{bij}W^{aij}_{fgh}
-\eta^{aij}_{fgh}W^{cde}_{bij}
\right)
\Bigg]\lambda^{cde}_{fgh},
\label{eq:df33-lambda3}
\end{align}

\begin{align}
&df^a_b(33,\lambda^{(4)}) \nonumber \\ 
=&
\sum_{cdefghij}\sum_k
\Bigg[
\frac{1}{16}
\left(
\eta^{acd}_{ghk}W^{efk}_{bij}
-\eta^{cdk}_{bgh}W^{aef}_{ijk}
\right)
\nonumber\\
&
+\frac{1}{24}
\Big(
\eta^{acd}_{bgk}W^{efk}_{hij}
-\eta^{ack}_{bgh}W^{def}_{ijk}
+\eta^{cdk}_{ghi}W^{aef}_{bjk}
-\eta^{cde}_{ghk}W^{afk}_{bij}
\Big)
\nonumber\\
&
+\frac{1}{36}
\left(
\eta^{ack}_{ghi}W^{def}_{bjk}
-\eta^{cde}_{bgk}W^{afk}_{hij}
\right)
\Bigg]\lambda^{cdef}_{ghij},
\label{eq:df33-lambda4}
\end{align}

\begin{align}
&df^a_b(33,\lambda^{(2)}\lambda^{(2)})
\nonumber\\
=&
\sum_{cdefghijk}
\Bigg[
\frac{1}{2}
\Big(
\eta^{acg}_{eik}W^{dhk}_{bfj}
-\eta^{cgk}_{bei}W^{adh}_{fjk}
+\eta^{ack}_{bei}W^{dgh}_{fjk}
-\eta^{acg}_{bek}W^{dhk}_{fij}
\nonumber\\
&
+\eta^{cdg}_{eik}W^{ahk}_{bfj}
-\eta^{cgk}_{eij}W^{adh}_{bfk}
\Big)
\nonumber\\
&
+\frac{1}{4}
\Big(
\eta^{ack}_{efi}W^{dgh}_{bjk}
-\eta^{cgh}_{bik}W^{adk}_{efj}
+\eta^{ack}_{eij}W^{dgh}_{bfk}
-\eta^{cdg}_{bik}W^{ahk}_{efj}
\nonumber\\
&
+\eta^{cdk}_{bei}W^{agh}_{fjk}
-\eta^{acg}_{ijk}W^{dhk}_{bef}
+\eta^{cgk}_{bef}W^{adh}_{ijk}
-\eta^{acd}_{eik}W^{ghk}_{bfj}
\Big)
\nonumber\\
&
+\frac{1}{8}
\Big(
\eta^{acd}_{bik}W^{ghk}_{efj}
-\eta^{ack}_{bij}W^{dgh}_{efk}
+\eta^{cdk}_{eij}W^{agh}_{bfk}
-\eta^{cdg}_{ijk}W^{ahk}_{bef}
\Big)
\nonumber\\
&
+\frac{1}{16}
\left(
\eta^{acd}_{ijk}W^{ghk}_{bef}
-\eta^{cdk}_{bij}W^{agh}_{efk}
\right)
\Bigg]
\lambda^{cd}_{ef}\lambda^{gh}_{ij}.
\label{eq:df33-lambda2lambda2}
\end{align}

\item \textbf{Two-body flow equation} 
The explicit expressions for all terms appearing on the right-hand side of Eq.~(\ref{eq:IMSRG3-2b}) are given below,
\begin{equation}
\begin{aligned}
    d\Gamma^{ab}_{cd}(12) 
    =&  \sum_{e} \hat{P}_{ab} (f^{a}_{e}\eta^{b e}_{c d} - \eta^{a}_{e}\Gamma^{b e}_{c d}) + \sum_{e} \hat{P}_{cd}(\eta^{e}_{c}\Gamma^{a b}_{d e} - f^{e}_{c}\eta^{a b}_{d e}),
\end{aligned}
\end{equation}

\begin{equation}
\begin{aligned}
    d\Gamma^{ab}_{cd}(13) = \sum_{ef}(n^{}_ {e} - n^{}_ {f})(\eta^{e}_ {f}W^{a b f}_ {c d e} - f^{e}_ {f}\eta^{a b f}_ {c d e} ),
\end{aligned}
\end{equation}

\begin{equation}
\begin{aligned}
    d\Gamma^{ab}_{cd}(22) &=  \sum_{ef} \Big[ \hat{P}_{ab}\hat{P}_{cd}(n_{e} - n_{f} )\eta^{a e}_ {c f}\Gamma^{b f}_ {d e} \\
    &+ \frac{1}{2} (n_e n_f - \bar{n}_e \bar{n}_f )(\eta^{e f}_ {c d}\Gamma^{a b}_ {e f} - \eta^{a b}_ {e f}\Gamma^{e f}_ {c d}) \Big].
\end{aligned}
\end{equation}
The $d\Gamma^{ab}_{cd}(23)$ term is composed of two terms,
\begin{equation}
    d\Gamma^{ab}_{cd}(23) = d\Gamma^{ab}_{cd}(23,\lambda^{(1)}) + d\Gamma^{ab}_{cd}(23,\lambda^{(2)}),
\end{equation}
which are obtained as
\begin{equation}
\begin{aligned}
   & d\Gamma^{ab}_{cd}(23,\lambda^{(1)})  \\
    &= \frac{1}{2}\sum_{efg}(n_e \bar{n}_f \bar{n}_g + \bar{n}_e n_f n_g)(\hat{P}_{cd}\eta^{fg}_{ce}W^{abe}_{dfg} - \hat{P}_{ab}\eta^{ae}_{fg}W^{bfg}_{cde}) \\
    &- \frac{1}{2} \sum_{efg}(n_e \bar{n}_f \bar{n}_g + \bar{n}_e n_f n_g)(\hat{P}_{cd}\Gamma^{fg}_{ce}\eta^{abe}_{dfg} - \hat{P}_{ab}\Gamma^{ae}_{fg}\eta^{bfg}_{cde}) ,
\end{aligned}
\end{equation}
and
\begin{equation}
\begin{aligned}
     & d\Gamma^{ab}_{cd}(23,\lambda^{(2)}) \\
    &=  \sum_{efghi} \Big[ \frac{1}{2}( \eta^{ab}_{gi} W^{efi}_{cdh} - \eta^{ei}_{cd} W^{abf}_{ghi}) \\
    &+ \frac{1}{4}( \hat{P}_{cd}\eta^{ef}_{ci} W^{abi}_{dgh} - \hat{P}_{ab}\eta^{ai}_{gh} W^{bef}_{cdi}) \\
    &+ \frac{1}{2}( \eta^{ef}_{gi} W^{abi}_{cdh} - \eta^{ei}_{gh} W^{abf}_{cdi}) \\
    &+ ( \hat{P}_{cd}\eta^{ei}_{cg} W^{abf}_{dhi} - \hat{P}_{ab}\eta^{ae}_{gi} W^{bfi}_{cdh}) \\
    &+ \frac{1}{2}\hat{P}_{ab}\hat{P}_{cd}( \eta^{ai}_{cg} W^{bef}_{dhi} - \eta^{ae}_{ci} W^{bfi}_{dgh})\Big] \lambda^{ef}_{gh} \\
    &- \sum_{efghi} \Big[ \frac{1}{2}( \Gamma^{ab}_{gi} \eta^{efi}_{cdh} - \Gamma^{ei}_{cd} \eta^{abf}_{ghi}) \\
    &+ \frac{1}{4}( \hat{P}_{cd}\Gamma^{ef}_{ci} \eta^{abi}_{dgh} - \hat{P}_{ab} \Gamma^{ai}_{gh} \eta^{bef}_{cdi}) \\
    &+ \frac{1}{2}( \Gamma^{ef}_{gi} \eta^{abi}_{cdh} - \Gamma^{ei}_{gh} \eta^{abf}_{cdi}) \\
    &+ ( \hat{P}_{cd}\Gamma^{ei}_{cg} \eta^{abf}_{dhi} -\hat{P}_{ab} \Gamma^{ae}_{gi} \eta^{bfi}_{cdh}) \\
    &+ \frac{1}{2}\hat{P}_{ab}\hat{P}_{cd}( \Gamma^{ai}_{cg} \eta^{bef}_{dhi} - \Gamma^{ae}_{ci} \eta^{bfi}_{dgh})\Big] \lambda^{ef}_{gh}.
\end{aligned}
\end{equation}
The $d\Gamma^{ab}_{cd}(33)$ term is composed of three terms,
\begin{equation}
    d\Gamma^{ab}_{cd}(33) = d\Gamma^{ab}_{cd}(33,\lambda^{(1)}) + d\Gamma^{ab}_{cd}(33,\lambda^{(2)}) + d\Gamma^{ab}_{cd}(33,\lambda^{(3)}),
\end{equation}
which are given by
\begin{equation}
\begin{aligned}
    & d\Gamma^{ab}_{cd}(33,\lambda^{(1)}) \\
    &= \sum_{efgh} \Big[ \frac{1}{4}\hat{P}_{ab}\hat{P}_{cd}(n_e n_f \bar{n}_g \bar{n}_h - \bar{n}_e \bar{n}_f n_g n_h)\eta^{aef}_{cgh}W^{bgh}_{def} \\
    &+ \frac{1}{6}(n_e \bar{n}_f \bar{n}_g \bar{n}_h - \bar{n}_e n_f n_g n_h)(\eta^{abe}_{fgh}W^{fgh}_{cde} - \eta^{fgh}_{cde}W^{abe}_{fgh}) \Big] ,
\end{aligned}
\end{equation}

\begin{equation}
\begin{aligned}
    d\Gamma^{ab}_{cd}(33,\lambda^{(2)}) 
    &= \sum_{efgh} \sum_{ij} \Big[ (n_i - n_j)\hat{P}_{ab}\hat{P}_{cd}\eta^{aei}_{cgj}W^{bfj}_{dhi} \\
    &- \frac{1}{2}(n_i - n_j)(\hat{P}_{cd}(\eta^{abi}_{cgj}W^{efj}_{dhi}+\eta^{efi}_{cgj}W^{abj}_{dhi}) \\
    &+\hat{P}_{ab}(\eta^{aei}_{cdj}W^{bfj}_{ghi}+\eta^{aei}_{ghj}W^{bfj}_{cdi})) \\
    &+ \frac{1}{4}(n_i - n_j)(\eta^{abi}_{ghj}W^{efj}_{cdi} + \eta^{efi}_{cdj}W^{abj}_{ghi}) \\
    &+ \frac{1}{2}(n_i n_j - \bar{n}_i \bar{n}_j)((\eta^{eij}_{cdg}W^{abf}_{hij} - \eta^{abe}_{gij}W^{fij}_{cdh})  \\
    &+ \frac{1}{2}(\hat{P}_{cd}\eta^{eij}_{cgh}W^{abf}_{dij} - \hat{P}_{ab}  \eta^{aef}_{gij}W^{bij}_{cdh}) \\
    &+ \frac{1}{2}(\hat{P}_{ab}\eta^{aij}_{cdg}W^{bef}_{hij}  -\hat{P}_{cd}\eta^{abe}_{cij}W^{fij}_{dgh}) \\
    &+ \frac{1}{4}\hat{P}_{ab}\hat{P}_{cd}(\eta^{aij}_{cgh}W^{bef}_{dij} - \eta^{aef}_{cij}W^{bij}_{dgh}) ) \Big] \lambda^{ef}_{gh} ,
\end{aligned}
\end{equation}

\begin{align}
&d\Gamma^{ab}_{cd}(33,\lambda^{(3)})
\nonumber\\
=&
\sum_{efghijk}
\Bigg[
\frac{1}{4}\hat{P}_{ab}\hat{P}_{cd}
\left(
\eta^{aek}_{chi}W^{bfg}_{djk}
-\eta^{aef}_{chk}W^{bgk}_{dij}
\right)
\nonumber\\
&+\frac{1}{4}
\Big(
\hat{P}_{ab}\eta^{aef}_{hik}W^{bgk}_{cdj}
-\hat{P}_{cd}\eta^{efk}_{chi}W^{abg}_{djk}
+\hat{P}_{ab}\eta^{aek}_{cdh}W^{bfg}_{ijk}
-\hat{P}_{cd}\eta^{abe}_{chk}W^{fgk}_{dij}
\Big)
\nonumber\\
&+\frac{1}{4}
\left(
\eta^{abe}_{hik}W^{fgk}_{cdj}
-\eta^{efk}_{cdh}W^{abg}_{ijk}
\right)
\nonumber\\
&+\frac{1}{12}
\Big(
\hat{P}_{ab}\eta^{aef}_{cdk}W^{bgk}_{hij}
-\hat{P}_{cd}\eta^{abk}_{chi}W^{efg}_{djk}
+\hat{P}_{ab}\eta^{aek}_{hij}W^{bfg}_{cdk}
-\hat{P}_{cd}\eta^{efg}_{chk}W^{abk}_{dij}
\Big)
\nonumber\\
&+\frac{1}{12}
\Big(
\eta^{abe}_{cdk}W^{fgk}_{hij}
-\eta^{abk}_{cdh}W^{efg}_{ijk}
+\eta^{efg}_{hik}W^{abk}_{cdj}
-\eta^{efk}_{hij}W^{abg}_{cdk}
\Big)
\nonumber\\
&+\frac{1}{36}
\left(
\eta^{efg}_{cdk}W^{abk}_{hij}
-\eta^{abk}_{hij}W^{efg}_{cdk}
\right)
\Bigg]\lambda^{efg}_{hij}.
\label{eq:dGamma33-lambda3}
\end{align}

\item \textbf{Three-body flow equation:}
 The explicit expressions for all terms appearing on the right-hand side of Eq.~(\ref{eq:IMSRG3-3b}) are given below, 
\begin{align}
    dW^{abc}_{def}(13) 
    &= \sum_{g} \hat{P}(a/bc)(\eta^{a}_ {g}W^{b c g}_ {d e f} - f^{a}_ {g}\eta^{b c g}_ {d e f}) \nonumber\\
    &+ \sum_{g} \hat{P}(d/ef)(f^{g}_ {d}\eta^{a b c}_ {e f g} - \eta^{g}_ {d}W^{a b c}_ {e f g}),
\end{align} 

\begin{equation}
\begin{aligned}
    dW^{abc}_{def}(22) = \sum_{g} \hat{P}(a/bc)\hat{P}(de/f)(\eta^{a g}_ {d e}\Gamma^{b c}_ {f g} - \eta^{bc}_ {f g}\Gamma^{a g}_ {de} ),
\end{aligned}
\end{equation}

\begin{align}
    dW^{abc}_{def}(23) 
    &= \sum_{gh} \Big[ (n_g - n_h)\hat{P}(a/bc)\hat{P}(d/ef)(\eta^{ag}_{dh} W^{bch}_{efg} - \Gamma^{ag}_{dh} \eta^{bch}_{efg})\nonumber\\
    &+ \frac{1}{2}(n_g n_h - \bar{n}_g \bar{n}_h)(\hat{P}(de/f)(\eta^{gh}_{de} W^{abc}_{fgh} - \Gamma^{gh}_{de} \eta^{abc}_{fgh}) \nonumber\\
    &- \hat{P}(ab/c)(\eta^{ab}_{gh} W^{cgh}_{def} - \Gamma^{ab}_{gh} \eta^{cgh}_{def})) \Big],
\end{align}

The $dW^{abc}_{def}(33)$ is composed of two terms
\begin{equation}
    dW^{abc}_{def}(33) = dW^{abc}_{def}(33,\lambda^{(1)}) + dW^{abc}_{def}(33,\lambda^{(2)}),
\end{equation}
which are given by
 \begin{align}
    & dW^{abc}_{def}(33,\lambda^{(1)}) \nonumber\\
    &= 
    \sum_{ghi} \Big[ \frac{1}{2}(n_g \bar{n}_h \bar{n}_i + \bar{n}_g n_h n_i)(\hat{P}(a/bc)\hat{P}(de/f)\eta^{ahi}_{deg}W^{bcg}_{fhi} \nonumber \\
    &- \hat{P}(ab/c)\hat{P}(d/ef)\eta^{abg}_{dhi}W^{chi}_{efg}) \nonumber\\ 
    &+ \frac{1}{6} (n_g n_h n_i + \bar{n}_g \bar{n}_h \bar{n}_i)(\eta^{abc}_{ghi}W^{ghi}_{def} - \eta^{ghi}_{def}W^{abc}_{ghi}) \Big] ,
\end{align} 
and 
\begin{align}
    dW^{abc}_{def}(33,\lambda^{(2)}) &= 
    \sum_{ghij} \sum_{k} \Big[ (\hat{P}(a/bc)\hat{P}(de/f)\eta^{a g k}_ {d e i}W^{b c h}_ {f j k} \nonumber\\
    &- \hat{P}(ab/c)\hat{P}(d/ef)\eta^{a b g}_ {d i k}W^{c h k}_ {e f j} )\nonumber \\
    &+ \frac{1}{2}(\hat{P}(ab/c)\hat{P}(de/f)(\eta^{a b k}_ {d e i}W^{c g h}_ {f j k} - \eta^{a b g}_ {d e k}W^{c h k}_ {f i j})\nonumber \\
    &+ \hat{P}(a/bc)\hat{P}(d/ef)(\eta^{a g h}_ {d i k}W^{b c k}_ {e f j}  - \eta^{a g k}_ {d i j}W^{b c h}_ {e f k})) \nonumber\\
    &+ \frac{1}{4}(\hat{P}(a/bc)\hat{P}(de/f)\eta^{a g h}_ {d e k}W^{b c k}_ {f i j} \nonumber\\
    &- \hat{P}(ab/c)\hat{P}(d/ef)\eta^{a b k}_ {d i j}W^{c g h}_ {e f k}) \nonumber\\
    &+ \frac{1}{2}(\hat{P}(d/ef)\eta^{a b c}_ {d i k}W^{g h k}_ {e f j} + \hat{P}(de/f)\eta^{g h k}_ {d e i}W^{a b c}_ {f j k} \nonumber\\
    &- \hat{P}(ab/c)\eta^{a b g}_ {i j k}W^{c h k}_ {d e f} - \hat{P}(a/bc)\eta^{a g k}_ {d e f}W^{b c h}_ {i j k}) \nonumber\\
    &+ \frac{1}{4}(\eta^{a b c}_ {i j k}W^{g h k}_ {d e f}    - \eta^{g h k}_ {d e f}W^{a b c}_ {i j k}) \Big] \lambda^{gh}_{ij}.
\end{align}

In the expressions above, $\hat{P}_{ab}\equiv 1-P_{ab}$ denotes the two-index antisymmetrization operator, where $P_{ab}$ is the permutation operator that interchanges the indices $a$ and $b$ throughout the expression on which it acts. For example,
\begin{equation}
P_{ab}\left(A^{ai}_{ck}B^{bj}_{dl}\right)
=
A^{bi}_{ck}B^{aj}_{dl}.
\end{equation}
Accordingly,
\begin{equation}
\hat{P}_{ab}\left(A^{ai}_{ck}B^{bj}_{dl}\right)
=
A^{ai}_{ck}B^{bj}_{dl}
-
A^{bi}_{ck}B^{aj}_{dl}.
\end{equation}

The notation $\hat{P}(a/bc)$ denotes the partial antisymmetrization of the index $a$ with respect to the index pair $b$ and $c$, and is defined as
\begin{equation}
\hat{P}(a/bc) \equiv \hat{P}(bc/a)
\equiv
1-P_{ab}-P_{ac}.
\end{equation}
This compact definition assumes that the expression on which $\hat{P}(a/bc)$ acts is already antisymmetric under the interchange $b\leftrightarrow c$.

\end{itemize}

\section{Comparison with several existing packages}
\label{sec:comparison_quantum_chemistry}
A number of symbolic-algebra frameworks have been developed to automate the derivation and implementation of quantum many-body theories,
particularly in quantum chemistry. The tensor contraction engine (TCE) \cite{Hirata2003TCE,Auer2006TCE} provides an end-to-end framework that combines symbolic Wick
contractions with tensor-contraction optimization and automatic
generation of parallel implementations for CI, CC, and many-body perturbation theory (MBPT) methods. The $p^\dagger q$ package \cite{Rubin2021pdaggerq,Liebenthal2025pdaggerq} provides a C++-accelerated Python interface for fermionic operator algebra and has recently been extended to optimize tensor contractions and generate executable Python and C++ implementations of coupled-cluster-type methods. {\tt SeQuant} \cite{Gaudel2026SeQuant,SeQuant2026Software} offers
a high-performance C++ framework for general symbolic tensor and
operator algebra, including Wick-theorem evaluation, tensor-network
canonicalization, correlated-reference support, and interfaces to numerical tensor backends such as {\tt TiledArray} \cite{Peng2016TiledArray}.
Similarly, {\tt SymGen}\cite{SymGen} is designed to generate tensor expressions for coupled-cluster and downfolding approaches and interfaces with the TAMM/ExaChem infrastructure for scalable numerical calculations
\cite{Mutlu2023TAMM,ExaChem,Bylaska2024Cloud,Bauman2026Downfolding}. Of
particular relevance to the present work, {\tt Wick\&D}~\cite{Evangelista2022Wickd} implements generalized
Wick algebra with respect to correlated reference states defined with an
arbitrary number of orbital subspaces and has been applied to multi-reference many-body theories.
 
The aim of \texttt{qcombo} is complementary to these packages. Rather than providing an end-to-end symbolic-to-numerical code-generation environment, \texttt{qcombo} is designed as a lightweight and transparent tool for the automated evaluation of commutators between general normal-ordered many-body operators. Correlated-reference algebra and the generalized Wick theorem are not unique to \texttt{qcombo}; for example, \texttt{Wick\&D} explicitly implements the Mukherjee--Kutzelnigg formalism, while \texttt{SeQuant} also provides correlated-reference functionality within a more general symbolic tensor framework. The advantage of \texttt{qcombo} therefore lies not in generalized normal ordering alone, but in its accessible and application-oriented combination of capabilities. Through a compact Python/\texttt{SymPy} interface, the same workflow treats both single-reference and correlated multi-reference states, retains the irreducible many-body density matrices explicitly, and allows the generated expressions to be filtered independently according to the body rank of the resulting operator and the rank of the density matrix involved. This organization facilitates the derivation, truncation, inspection, and validation of high-rank multi-reference equations. In nuclear many-body applications,  there are additional algebraic complexity through the spin, isospin, and tensor structure of the interactions, as well as the presence of three-nucleon forces. \texttt{qcombo} can generate input directly for the \texttt{AMC} package \cite{Tichai2020amc}, allowing the $M$-scheme expressions produced by \texttt{qcombo} to be subsequently transformed into angular-momentum-coupled $J$-scheme expressions.  Such a transformation is essential for exploiting rotational symmetry and obtaining practical equations in spherical single-particle bases, including for non-scalar observables and transition operators.

Moreover, the scope of \texttt{qcombo} is not restricted to the IMSRG. Because its central functionality is the evaluation of commutators between general normal-ordered operators, the combined \texttt{qcombo}--\texttt{AMC} workflow can also support commutator-based developments in nuclear coupled-cluster and equation-of-motion methods, Green's-function techniques, and the consistent similarity transformation of observables and transition operators. The derivation of the complete MR-IMSRG(3) equations presented in this work should therefore be viewed as a demanding representative application of the general framework rather than as a limitation on its range of applicability.


\end{document}